\documentclass[pre,aps,twocolumn,superscriptaddress]{revtex4-1}

\usepackage{graphicx}
\usepackage{amssymb,amsfonts,amsmath}
\usepackage{color}
\usepackage{ulem}
\usepackage[hidelinks]{hyperref}
\usepackage{bbm}
\usepackage{bbold}

\usepackage{tikz}
\usetikzlibrary{shapes}
\usetikzlibrary{patterns}
\usetikzlibrary{angles, quotes}

\newcommand{\rv}{{\mathbf r}}

\newcommand{\Tr}{{\rm Tr}\,}
\newcommand{\e}{{\rm e}}

\newcommand{\pv}{{\bf p}}

\newcommand{\Fv}{{\bf F}}

\newcommand{\msphantom}[1]{$\ldots$}

\newcommand{\eqr}[1]{Eq.~\eqref{#1}}

\newcommand{\mydelete}[1]{{}}

\newcommand{\rmint}{{\rm int}}

\newcommand{\rmexc}{{\rm exc}}
\newcommand{\rmext}{{\rm ext}}
\newcommand{\rmeq}{{\rm eq}}
\newcommand{\rmid}{{\rm id}}

\newcommand{\cov}{{\rm cov}}

\newcommand{\Sv}{{\bf S}}

\begin{document}

\title{Why hyperdensity functionals describe any equilibrium observable}

\author{Florian Samm\"uller}
\affiliation{Theoretische Physik II, Physikalisches Institut, 
  Universit{\"a}t Bayreuth, D-95447 Bayreuth, Germany}

\author{Matthias Schmidt}
\affiliation{Theoretische Physik II, Physikalisches Institut, 
  Universit{\"a}t Bayreuth, D-95447 Bayreuth, Germany}
\email{Matthias.Schmidt@uni-bayreuth.de}

\date{14 October 2024, revised version: 23 November 2024}

\begin{abstract}
  We give an introductory account of the recent hyperdensity
  functional theory for the equilibrium statistical mechanics of soft
  matter systems [F.~Samm\"uller {\it et
      al.,}~\href{https://doi.org/10.1103/PhysRevLett.133.098201}
    {Phys.\ Rev.\ Lett.\ {\bf 133}, 098201 (2024)}].  Hyperdensity
  functionals give access to the behaviour of arbitrary thermal
  observables in spatially inhomogeneous equilibrium many-body
  systems. The approach is based on classical density functional
  theory applied to an extended ensemble using standard functional
  techniques. The associated formally exact generalized Mermin-Evans
  functional relationships can be represented accurately by neural
  functionals. These neural networks are trained via simulation-based
  supervised machine learning and they allow one to carry out
  efficient functional calculus using automatic differentiation and
  numerical functional line integration.  Exact sum rules, including
  hard wall contact theorems and hyperfluctuation Ornstein-Zernike
  equations, interrelate the different correlation functions.  We lay
  out close connections to hyperforce correlation sum rules
  [S.~Robitschko {\it et
      al.,}~\href{https://doi.org/10.1038/s42005-024-01568-y}{Commun. Phys.
      {\bf 7}, 103 (2024)}] that arise from statistical mechanical
  gauge invariance [J.~M\"uller {\it et
      al.,}~\href{https://doi.org/10.1103/PhysRevLett.133.217101}{Phys. Rev. Lett. {\bf
        133}, 217101 (2024)}].  Further quantitative measures of
  collective self-organization are provided by hyperdirect correlation
  functionals and spatially resolved hyperfluctuation profiles.  The
  theory facilitates to gain deep insight into the inherent
  structuring mechanisms that govern the behaviour of both simple and
  complex order parameters in coupled many-body systems.
\end{abstract}

\maketitle

\section{Introduction}
\label{SECintroduction}

The Mermin-Evans theorem \cite{mermin1965, evans1979, evans1992,
  evans2016, hansen2013, schmidt2022rmp} of density functional theory
provides the finite temperature and classical mechanical
generalization of the pivotal Hohenberg-Kohn
proof~\cite{hohenberg1964, kohn1999nobel}.  These theorems enable a
full representation of the equilibrium statistical mechanics of
particle-based systems via systematically constructed functional
dependencies.  In particular the eponymous one-body density profile
$\rho(\rv)$, where $\rv$ denotes spatial position, plays a leading
role as a variational variable. Higher-body correlation functions are
accessible both via the Ornstein-Zernike relationship
\cite{ornstein1914, hansen2013} and Percus' test particle route
\cite{percus1962, hansen2013} to represent the full correlated physics
that emerges from the underlying coupling of the individual particles.
Full thermodynamic information, including bulk and interfacial
contributions, is thereby accessible.

In typical applications one is interested in situations where spatial
inhomogeneity is induced by the action of an external potential
$V_\rmext(\rv)$.  Appropriate choices of the form of $V_\rmext(\rv)$
allow one to model a broad class of different types of external
influence that is exerted onto a system, including walls, confinement,
the behaviour around solutes, {\it etc.}  The microscopic degrees of
freedom of the system are thereby coupled through an interparticle
interaction potential $u(\rv^N)$ that depends on all coordinates
$\rv^N=\rv_1,\ldots,\rv_N$ of the $N$ particles. Typically the
thermodynamic statepoint is given via prescribing values of the
chemical potential $\mu$ and of the absolute temperature~$T$.

From an elementary statistical mechanical point of view it is in
principle a standard task to predict the density profile $\rho(\rv)$,
which in the classical realm is straightforward to accomplish in
computer simulations. One needs to average the density operator
$\hat\rho(\rv)= \sum_i \delta(\rv-\rv_i)$, where the sum over $i$ runs
over all $N$ particles,
and $\delta(\cdot)$ indicates the Dirac distribution. The thermal
ensemble average is indicated by $\langle\cdot\rangle$ and it can
readily be realized via importance sampling on the basis of
e.g.\ grand canonical Monte Carlo simulations \cite{frenkel2023book,
  wilding2001, brukhno2021dlmonte}.  Thereby powerful histogram
\cite{wilding2001}, force sampling \cite{frenkel2023book, borgis2013,
  delasheras2018forceSampling, coles2019, coles2021, rotenberg2020,
  mangaud2020, coles2023revelsMD,renner2023torqueSampling} and mapped
averaging \cite{moustafa2015, schultz2016, moustafa2017jctp,
  moustafa2017prb, schultz2018, purohit2018, moustafa2019,
  purohit2020, moustafa2022, lin2018, trokhymchuk2019, schultz2019,
  purohit2019} techniques are available.

Given this quite pedestrian status it can seem surprising that the
density profile $\rho(\rv)=\langle \hat\rho(\rv)\rangle$ is of
Nobel-prize winning format in the quantum
realm~\cite{kohn1999nobel}. The situation can seem even more
perplexing as $\rho(\rv)$ is the only relevant variational variable in
density functional theory, which hence seemingly lacks any explicit
occurrence of two- and higher-body correlation functions. Using any
further problem-specific, tailor-made order parameters is also quite
alien to the framework.

That these apparent deficiencies can all be remedied by the mere
inversion of the functional map, i.e., realizing and establishing
$\rho(\rv)\to V_\rmext(\rv)$, can seem mysterious. Possibly at the
center of the mystery is the density functionalists' credo that
``everything is a density functional'', which poignantly expresses
that, for given interparticle interaction potential $u(\rv^N)$, from
knowledge of $\rho(\rv)$ the Hamiltonian itself can be
reconstructed. Once the Hamiltonian is known along with the
thermodynamic conditions, {\it any} equilibrium property, no matter
how complex or intricate, is known in principle and has thus become a
density functional. This formal structure can certainly seem
surprising and we here wish to lay out its concrete consequences and
route to practical implementation.

Functional relationships have acquired new and compelling relevance in
light of the recent neural functional theory
\cite{sammueller2023neural, sammueller2023whyNeural,
  sammueller2023neuralTutorial, sammueller2024pairmatching,
  sammueller2024attraction, sammueller2024hyperDFT,
  delasheras2023perspective, zimmermann2024ml, bui2024neuralrpm,
  kampa2024meta}. This hybrid approach utilizes many-body simulations
to generate data for supervised training of an artificial neural
network, which then acts as a neural functional. A number of features
set this approach apart from more generic machine-learning methods
\cite{carrasquilla2017, bedolla2021, chertenkov2023, arnold2024}, from
physics-informed machine learning in liquid state integral equation
theory \cite{carvalho2022ml, chen2024ml, wu2023review}, as well as
from other uses of machine learning in classical \cite{lin2019ml,
  lin2020ml, cats2022ml, yatsyshin2022, malpica-morales2023,
  dijkman2024ml, kelley2024ml, simon2023mlPatchy, simon2024patchy,
  stierle2024autodiff, yang2024} and in quantum density functional
theory \cite{nagai2018, jschmidt2018, zhou2019, nagai2020, li2021prl,
  li2022natcompsci, pederson2022, gedeon2022}.  As is argued in
Refs.~\cite{sammueller2023neural, sammueller2023whyNeural,
  sammueller2023neuralTutorial, sammueller2024pairmatching,
  sammueller2024attraction, sammueller2024hyperDFT,
  delasheras2023perspective, zimmermann2024ml, bui2024neuralrpm,
  kampa2024meta}, the neural functional theory constitutes a genuine
theoretical framework that permits one to carry out deep functional
calculus and to obtain a very complete picture of the physics under
investigation.

We summarize several key features of neural functionals. i)~Learning
of the relationship between input and output data pairs is based on a
rigorous mathematical relationship, which is known from first
principles to exist and to be unique, see the discussion given in
Ref.~\cite{sammueller2024pairmatching}. ii) Local learning
\cite{sammueller2023neural,
  sammueller2023whyNeural,sammueller2023neuralTutorial,
  sammueller2024pairmatching} facilitates data-efficient training and
subsequent ``beyond-the-box'' application of the resulting neural
functional to challenging multi-scale problems. iii) Efficient
implementation of functional calculus is provided by automatic
functional differentiation \cite{baydin2018autodiff} and fast
numerical functional integration. iv)~The neural representations of
both direct correlation functionals and of free energy functionals are
accurate and performant. v) The internal consistencies of a neural
functional can be tested via numerical evaluation of a wide variety of
exact statistical mechanical sum rules.

The body of literature addressing statistical mechanical sum rules,
i.e., exact identities that hold universally, is both large and
diverse; see e.g.\ Refs.~\cite{baus1984, evans1990,
  henderson1992}. The recent thermal Noether invariance theory
provides a systematic approach for both the derivation and the
classification of sum rules \cite{hermann2021noether,
  hermann2022topicalReview, hermann2022variance, hermann2022quantum,
  tschopp2022forceDFT, sammueller2023whatIsLiquid,
  hermann2023whatIsLiquid, robitschko2024any, mueller2024gauge,
  mueller2024whygauge}. The thermal Noether invariance is thereby
inherent in the very foundations of the statistical mechanics and
constitutes a gauge transformation for statistical mechanical
microstates \cite{mueller2024gauge, mueller2024whygauge}; see the very
recent Viewpoint given by Rotenberg \cite{rotenberg2024spotted}. The
approach is free of simplifying assumptions and
approximations. Technically the symmetry is an invariance of the phase
space integral under specific transformation operations of the phase
space variables. The Noether framework not only generates exact
identities, but it also acts as a construction device to generate the
specific correlation functions for which these identities hold.  The
correlation functions range from standard density-based statistical
mechanical correlations to force-based observables, which were shown
to shine new light on liquid structure even in
bulk~\cite{sammueller2023whatIsLiquid, hermann2023whatIsLiquid}.

In a remarkable contribution Hirschfelder \cite{hirschfelder1960}
generalized the standard virial theorm \cite{hansen2013}, which dates
back to Clausius and the very origins of thermodynamics, to include an
arbitrary observable $\hat A$, as represented by a phase space
function in the present classical context. The hyperforce theory of
Ref.~\cite{robitschko2024any} performs a similar generalization of the
force-balance relationship, as expressed in local form via a hierarchy
of equations due to Yvon~\cite{yvon1935} and Born and Green
\cite{born1946}.  Hyperforce sum rules hold both globally as well as
locally resolved in position and they couple in specific ways the
general observable $\hat A$ with the fundamental degrees of freedom of
the system via force- and density-based correlation functions.

We recall that the one-body force balance equation integrates itself
very naturally into density functional theory, where it is recovered
as the spatial gradient of the fundamental Euler-Lagrange equation
\cite{evans1979, schmidt2022rmp, tschopp2022forceDFT}.  An analogous
correspondence exists for the hyperforce framework
\cite{robitschko2024any, mueller2024gauge, mueller2024whygauge}, as
this is mirrored and complemented by the corresponding hyperdensity
functional theory \cite{sammueller2024hyperDFT}.  This approach
facilitates the explicit construction of the mean of the observable
$\hat A$ as a density functional, i.e., $A[\rho]$, where the brackets
indicate the functional relationship. Here we give an introductory and
extended account of the hyperdensity functional theory
\cite{sammueller2024hyperDFT}, including a detailed discussion of its
relationship with the hyperforce sum rules of
Ref.~\cite{robitschko2024any} and the underlying gauge invariance
concept \cite{mueller2024gauge, mueller2024whygauge,
  rotenberg2024spotted}.

Briefly, hyperdensity functional theory \cite{sammueller2024hyperDFT}
allows one to address the equilibrium behaviour of complex order
parameters $\hat A$ in a tightly integrated and concrete statistical
mechanical framework.  The equilibrium average of $\hat A$, expressed
as a density functional $A[\rho]$, is associated with a hyperdirect
correlation functional $c_A(\rv;[\rho])$ and a hyperfluctuation
profile $\chi_A(\rv)$, both of which are specific to the form of $\hat
A$.  The one-body hyperdirect correlation functional $c_A(\rv;[\rho])$
plays a role similar to that of the one-body direct correlation
functional $c_1(\rv;[\rho])$ of standard density functional
theory. The hyperfluctuation profile $\chi_A(\rv)$ can be viewed as a
generalization of thermodynamic fluctuation profiles
\cite{evans2015jpcm, evans2019pnas, coe2022prl, coe2022pre, coe2023,
  eckert2020, eckert2023fluctuation, wilding2024}, such as the
well-studied local compressibility $\chi_\mu(\rv)$
\cite{evans2015jpcm, evans2019pnas, coe2022prl, wilding2024}.

\begin{figure}[htb!]
  \includegraphics[width=0.45\textwidth]{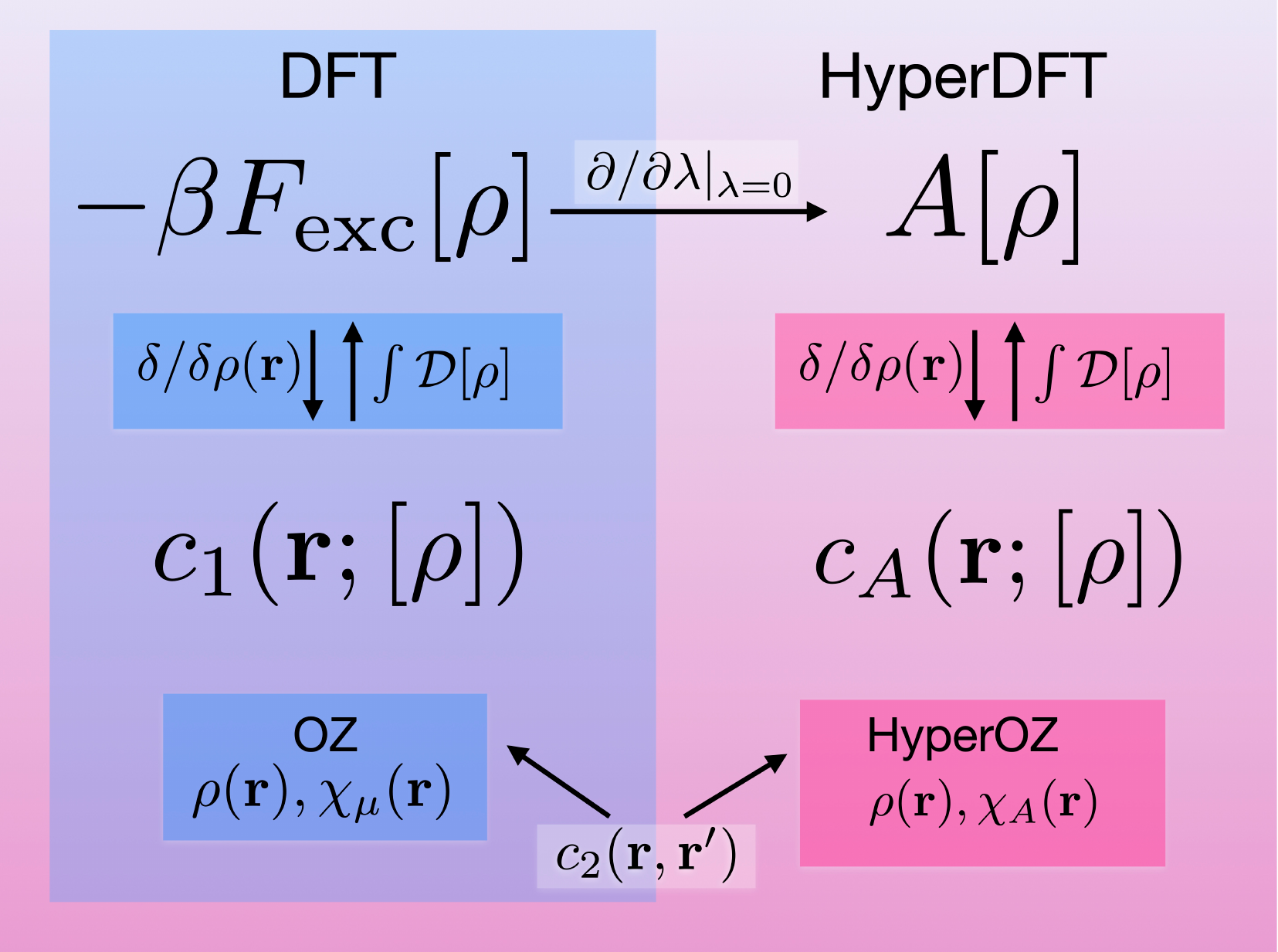}
    \caption{Functional relationships of classical density functional
      theory (left column) and the hyperdensity functional
      generalization (right column). The scaled excess free energy
      functional $-\beta F_\rmexc[\rho]$ is related to the one-body
      direct correlation functional $c_1(\rv;[\rho])$ via functional
      differentiation, $\delta/\delta\rho(\rv)$, and, inversely, by
      functional integration, $\int {\cal D}[\rho]$. Analogous
      relationships connect the hyperdensity functional $A[\rho]$ that
      expresses the mean of any given observable $\hat A$ and the
      hyperdirect correlation functional $c_A(\rv;[\rho])$. The local
      compressibility $\chi_\mu(\rv)$ and the hyperfluctuation profile
      $\chi_A(\rv)$ play analogous roles and they are respectively
      connected via Ornstein-Zernike and hyper-Ornstein-Zernike
      relations, which both feature the two-body direct correlation
      functional $c_2(\rv,\rv';[\rho])$. An extended ensemble allows
      differentiating with respect to the coupling parameter,
      $\partial/\partial\lambda|_{\lambda=0}$.
    \label{FIG1}
   }
\end{figure}

Together with the standard two-body direct correlation functional
$c_2(\rv,\rv';[\rho])$, the correlation functions $c_A(\rv;[\rho])$
and $\chi_A(\rv)$ are connected via an exact hyper-Ornstein-Zernike
relation~\cite{sammueller2024hyperDFT}, which generalizes the standard
inhomogeneous two-body Ornstein-Zernike relation \cite{hansen2013,
  evans1979, schmidt2022rmp} to general observables $\hat A$.  The
hyper-Ornstein-Zernike relation is of relatively simple one-body form
(two-body functions feature only inside of a spatial integral) and the
mathematical structure is akin to the one-body
fluctuation-Ornstein-Zernike relationships~\cite{eckert2020,
  eckert2023fluctuation}. The fluctuation-Ornstein-Zernike equation
for the local compressiblity $\chi_\mu(\rv)$ \cite{eckert2020,
  eckert2023fluctuation} was shown to deliver efficiently accurate
results in demanding drying situations using neural density functional
methods together with automatic differentiation
\cite{sammueller2024attraction}. An illustration of the relationship
between the key quantities is shown in Fig.~\ref{FIG1}.

In exemplary applications to concrete systems, the hyperdensity
functional was used for the investigation of clustering properties of
standard model fluids. The behaviour of one-dimensional hard rods was
compared to that of square-well attractive rods and of hard sphere
fluids in three dimensions. Our definition of the clustering follows a
standard procedure (see, e.g., Ref.~\cite{sammueller2023gel}). One
starts with a bonding criterion that declares two particles as bonded
provided that their mutual distance is smaller than some cutoff (taken
to be $1.2\sigma$ with $\sigma$ indicating the particle
size). Although the bonding criterion itself is a two-body function,
the resulting graph structure of bonded particles is non-trivial with
naive implementations requiring iterated passes over all particles,
but more efficient algorithms can be used \cite{sammueller2023gel,
  sammueller2024hyperDFT}. In a second step the number of particles in
each cluster is counted and the largest such number is searched for,
which then is taken as the observable $\hat A$. Each microstate gives
a unique value of the size of the largest cluster $\hat A$, hence
$\hat A$ is indeed a phase space function, as is required in the
hyperdensity functional framework. However, in contrast to standard
phase space functions such as the Hamiltonian {\it etc.}, no closed
expression for $\hat A$ is available and hence $\hat A$ constitutes an
{\it algorithmically defined} observable. Nevertheless, as we lay out,
the thermal mean $A=\langle \hat A \rangle$ is a unique hyperdensity
functional.

The paper is structured as follows. In Sec.~\ref{SEChyperDFT} we
describe the hyperdensity functional theory of
Ref.~\cite{sammueller2024hyperDFT}. Specifically, the extended
ensemble for the treatment of general observables $\hat A$ is
described in Sec.~\ref{SECextendedEnsemble}.  The average $\langle
\hat A \rangle$ and the local hyperfluctuation profile $\chi_A(\rv)$
are introduced in Sec.~\ref{SECgeneralObservables}. The Mermin-Evans
minimization principle of classical density functional theory, as is
at the heart of the hyperdensity generalization, is described in
Sec.~\ref{SECminimizationPrinciple}. A brief account of Levy's
constrained search method for the construction of the intrinsic free
energy density density functional is given in Sec.~\ref{SEClevy}.  The
hyper-Ornstein-Zernike relation is derived in
Sec.~\ref{SEChyperOZgeneral}. General observables are expressed as
hyperdensity functionals in Sec.~\ref{SEChyperDensityFunctionals}.
The wall hypercontact theorem is presented in
Sec.~\ref{SECcontactTheorem}.

In Sec.~\ref{SEChyperforce} we relate the hyperdensity functional
theory to the hyperforce theory of Ref.~\cite{robitschko2024any} as it
arises from statistical mechanical gauge invariance
\cite{mueller2024gauge, mueller2024whygauge}. In particular we derive
in Sec.~\ref{SEClocalHyperForceBalance} the exact one-body hyperforce
balance relationship \cite{robitschko2024any} from the extended
ensemble, which is an alternative both to the Noether functional
invariance method~\cite{robitschko2024any} and to the phase space
operator approach~\cite{mueller2024gauge, mueller2024whygauge}. In
Sec.~\ref{SEChyperforceConnections} we present exact sum rules that
connect the hyperforce correlation functions with the hyperdensity
functionals.  In Sec.~\ref{SECclassifyingObservables} we treat
specific simple cases of choice of the observable $\hat A$ including
one-body (Sec.~\ref{SEConeBodyObervables}) and two-body forms
(Sec.~\ref{SECtwoBodyObervables}).  We describe applications in
Sec.~\ref{SECapplications}, including a description of the workflow
that is required in the hyperdensity functional studies in
Sec.~\ref{SECworkflow}, the training of neural hyperdensity
functionals in Sec.~\ref{SECtraining}, and a model application to
clustering of hard spheres in Sec.~\ref{SECapplicationHardSpheres}.
We give our conclusions and an outlook in Sec.~\ref{SECconclusions}.

\section{Hyperdensity functional theory}
\label{SEChyperDFT}

\subsection{Extended statistical ensemble}
\label{SECextendedEnsemble}

We start by introducing the generalized equilibrium grand ensemble
used in Ref.~\cite{sammueller2024hyperDFT} as a basis for the
statistical mechanics. The extended ensemble facilitates to
incorporate the statistical behaviour of a given observable $\hat A$,
which can be arbitrarily chosen, into the density functional
framework. The use of an extended ensemble is a standard means of
generalization, see e.g.\ Refs.~\cite{zwanzig2001, anero2013}. In
their remarkable contribution Anero {\it et al.}~\cite{anero2013}
formulated a functional thermodynamics as a generalization of dynamic
density functional theory \cite{evans1979, marconi1999, espanol2009,
  schmidt2022rmp, delasheras2023perspective} to non-isothermal
situations.

Here we are ultimately interested in the equilibrium properties of
systems with Hamiltonians of the following standard form:
\begin{align}
  H=\sum_i \frac{\pv_i^2}{2m} + u(\rv^N) + \sum_i V_\rmext(\rv_i),
  \label{EQhamiltonian}
\end{align}
where $\pv_i$ denotes the momentum of particle $i$, the variable $m$
indicates the particle mass, $u(\rv^N)$ is the interparticle
interaction potential as a function of all position coordinates
$\rv^N=\rv_1,\ldots, \rv_N$, the external one-body potential is
$V_\rmext(\rv)$, and the sums over $i$ run over all $N$ particles,
such that $i=1,\ldots,N$.

We wish to address the properties of a given phase space function
$\hat A(\rv^N)$ which represents a physically relevant observable,
such as an order parameter that characterizes the behaviour of the
system specified by the Hamiltonian \eqref{EQhamiltonian}. We consider
the system to be coupled to both a heat bath with absolute
temperature~$T$ and to a particle reservoir that sets the chemical
potential $\mu$.  Hence the corresponding standard grand canonical
Boltzmann factor is $\e^{-\beta(H-\mu N)}$, where $\beta=1/(k_BT)$
with $k_B$ denoting the Boltzmann constant.  According to standard
procedure the partition sum is $\Tr \e^{-\beta(H-\mu N)}$ and the
grand potential is $-k_BT\ln \Tr \e^{-\beta (H-\mu N)}$.  Here the
classical ``trace'' operation is defined in the standard way as
$\Tr=\sum_{N=0}^\infty (N!h^{dN})^{-1}\int d\rv^N d\pv^N$, where $h$
denotes the Planck constant, $d$ is the space dimension, and $\int
d\rv^N d\pv^N$ indicates the phase space integral over all position
and momentum variables.

In order to address the statistical behaviour of a given observable
$\hat A(\rv^N)$ we consider an extended setup, which can be described
in two equivalent ways. First, we consider the Hamiltonian
\eqref{EQhamiltonian} to be unchanged but extend the statistical
ensemble itself, such that its generalized Boltzmann factor is
$\e^{-\beta (H-\mu N) + \lambda\hat A}$, where the coupling parameter
$\lambda$ regulates the degree of influence of $\hat A$ on the
probability distribution function. The corresponding normalization
factor is the extended partition sum given by $\Xi = \Tr
\e^{-\beta(H-\mu N)+\lambda\hat A}$. Averages in the extended ensemble
are then obtained as $\langle \cdot \rangle = \Tr \cdot
\e^{-\beta(H-\mu N)+\lambda\hat A}/\Xi$ and the extended grand
potential (or extended grand canonical free energy) is
$\Omega=-k_BT\ln\Xi$ and this depends per construction parametrically
on $\lambda$.  Despite the generalization, we remain ultimately
interested in the limit $\lambda\to 0$, which however is taken
typically only after carrying out the appropriate derivatives.

The second route leads to the identical statistical physics and it is
based on modifying the Hamiltonian itself rather than merely the
statistical ensemble used for the description. Here one defines an
extended Hamiltonian
\begin{align}
  H_A = H - \lambda \hat A/\beta,
  \label{EQhamiltonianA}
\end{align}
where $H$ is the original Hamiltonian according to \eqr{EQhamiltonian}
and the observable $\hat A$ is hence considered to actually contribute
to the interactions that define the system. Again the parameter
$\lambda$ tunes the strength of these now extended interactions and as
before the physical units of $\lambda$ are those of the inverse to
$\hat A(\rv^N)$ such that the second term in \eqr{EQhamiltonianA} has
units of energy, which arise from $1/\beta=k_BT$.

In this second route the extended Hamiltonian \eqref{EQhamiltonianA}
is fed in a straightforward way into the standard grand ensemble
machinery. Hence the Boltzmann factor is $\e^{-\beta(H_A-
  \mu N)}=\e^{-\beta (H-\lambda \hat A/\beta-\mu N)} = \e^{-\beta (H-\mu
  N)+\lambda\hat A}$, where in the first step we have replaced $H_A$
according to \eqr{EQhamiltonianA} and in the second step have
re-ordered the terms. The resulting expression is identical to the
Boltzmann factor of the extended ensemble according to the above first
route.  Following the standard procedure the resulting partition sum
is then $\Xi=\Tr \e^{-\beta(H_A-\mu N)}$, the thermal equilibrium
average is defined as $\langle\cdot\rangle=\Tr\e^{-\beta(H_A-\mu
  N)}/\Xi$, and the grand potential is $\Omega=-k_BT\ln\Xi$. These
expressions are all equivalent to their respective above counterparts
from the first route via the ensemble extension.

\subsection{Observables beyond the density profile}
\label{SECgeneralObservables}

As laid out in the introduction, density functional theory assigns a
special role to the equilibrium one-body density distribution
$\rho(\rv)$. When expressed as a thermal average the density profile
is simply given as
\begin{align}
 \rho(\rv)&=\langle\hat\rho(\rv)\rangle.
 \label{EQrhoAsAverage}
\end{align}
The one-body density ``operator'' (phase space function)
$\hat\rho(\rv)$ has the standard form
\begin{align}
  \hat\rho(\rv)=\sum_i\delta(\rv-\rv_i),
  \label{EQdensityOperator}
\end{align}
with $\delta(\cdot)$ denoting the Dirac distribution in $d$
dimensions. We consider the average in \eqr{EQrhoAsAverage} to be
taken over the extended ensemble. Hence this is specific to the form
of $\hat A(\rv^N)$ and it depends parametrically on the value of the
coupling parameter $\lambda$.
Taking $\lambda\to 0$ restores the original Hamiltonian, $H_A\to H$,
and thus $\rho(\rv)$ reduces to the density profile that corresponds
to the ``real'' Hamiltonian $H$.

Besides thermodynamical quantities, such as the free energy and the
pressure, density functional theory encompasses in principle two- and
higher-body correlation functions, as defined in generalization of
\eqr{EQrhoAsAverage} via the Ornstein-Zernike route; we sketch the
mathematical structure below in Sec.~\ref{SEChyperOZgeneral}. For
systems that interact only with pairwise interparticle forces this
allows one to express the locally resolved force density within a
force-based formulation of density functional theory
\cite{tschopp2022forceDFT, sammueller2022forceDFT}. Furthermore,
hyperforce sum rules can be exploited to obtain a range of additional
correlation functions from standard density functional calculations;
we refer the Reader to the conclusions of
Ref.~\cite{robitschko2024any} for a discussion of these opportunities.

Here we generalize further and hence are interested in the behaviour
of a given observable $\hat A(\rv^N)$.  We first consider its mean
value
\begin{align}
  A &= \langle \hat A \rangle,
  \label{EQmeanAasAverage}
\end{align}
which is a global quantity provided that $\hat A(\rv^N)$ itself
carries no further dependence on position. Alternatively, in case
$\hat A(\rv^N;\rv,\rv',\ldots)$ carries further dependence on generic
position variables $\rv,\rv',\ldots$ then the mean
$A(\rv,\rv',\ldots)$ will be a spatially resolved (correlation)
function.

In order to address the average \eqref{EQmeanAasAverage} via the
extended ensemble of Sec.~\ref{SECextendedEnsemble}, we first revert
to the standard mechanism of generating averages via parametric
derivatives. In the present case we have
\begin{align}
  A &= -\frac{\partial \beta\Omega}{\partial \lambda},
  \label{EQmeanAparametricDerivative}
\end{align}
where we recall the definition of the extended grand potential
$\Omega=-k_BT \ln \Xi$ with the extended partition sum $\Xi=\Tr
\e^{-\beta(H-\mu N)+\lambda \hat A}$. Both the form of the external
potential $V_\rmext(\rv)$ as well as the statepoint $\mu,T$ are kept
fixed when differentiating with respect to $\lambda$ in
\eqr{EQmeanAparametricDerivative}. The validity of this relationship
can be seen from standard parametric
differentiation~\cite{hansen2013}: $A=\partial \ln \Xi/\partial
\lambda=\Xi^{-1} \Tr \partial \e^{-\beta(H-\mu N)+\lambda\hat A}
/\partial \lambda=\Xi^{-1}\Tr\e^{-\beta(H-\mu N)+\lambda\hat A}\hat
A=\langle\hat A\rangle$.

In order to also systematically incorporate locally resolved
fluctuations of $\hat A$ we construct a one-body fluctuation profile
$\chi_A(\rv)$. The inspiration stems from the local compressiblity
\cite{evans2015jpcm, evans2019pnas, coe2022prl, coe2022pre, coe2023,
  eckert2020, eckert2023fluctuation, wilding2024} and the local
thermal susceptibility \cite{eckert2020, coe2022pre,
  eckert2023fluctuation}, which respectively correlate the
fluctuations of the local density with the total number of particles
and the entropy.

In the present case of a general observable $\hat A$, the
hyperfluctuation profile is obtained as the covariance of the
considered observable with the position-dependent density operator
\eqref{EQdensityOperator} according to
\begin{align}
  \chi_A(\rv) &= \cov(\hat\rho(\rv), \hat A)
  \label{EQchiAsCovariance}\\
  &= \langle\hat\rho(\rv)\hat A\rangle - \rho(\rv)A,
  \label{EQchiAsCovarianceExplicit}
\end{align}
where \eqr{EQchiAsCovarianceExplicit} is the explicit form of the
covariance in \eqr{EQchiAsCovariance}. The covariance of two general
phase space functions $\hat X$ and $\hat Y$ is $\cov(\hat X, \hat
Y)=\langle \hat X \hat Y\rangle - \langle \hat X \rangle \langle \hat
Y \rangle$.  We recall that by construction for cases where the mean
product factorizes according to $\langle\hat X\hat Y\rangle=\langle
\hat X \rangle \langle \hat Y \rangle$, then by its very definition
the covariance vanishes, $\cov(\hat X, \hat Y)=0$. Hence nonvanishing
covariance of two observables is a measure of the degree of deviation
from an idealized factorization behaviour. Using the generic
definition of the covariance in \eqr{EQchiAsCovariance} gives the
hyperfluctuation profile $\chi_A(\rv)$ in the more explicit form
\eqref{EQchiAsCovarianceExplicit}, where the density profile
$\rho(\rv)$ is given by \eqr{EQrhoAsAverage} and the mean $A$ is given
by \eqr{EQmeanAasAverage} and, equivalently, by
\eqr{EQmeanAparametricDerivative}.

While defining the mean $A$ via \eqr{EQmeanAasAverage} is entirely
standard, the definition of the hyperfluctuation profile $\chi_A(\rv)$
via \eqr{EQchiAsCovariance} introduces position-dependence by
correlating the presence of a particle at a specific position~$\rv$
with the overall value of $\hat A(\rv^N)$. The hyperfluctuation
profile is hence not a merely ``local'' version of the order
parameter~$\hat A(\rv^N)$ in the sense that a space integral of the
local version gives the global version. Rather the position integral
gives $\int d\rv \chi_A(\rv)=\int d\rv \cov(\hat\rho(\rv),\hat
A)=\cov(N,\hat A)$, where the latter equality holds due to $\int d\rv
\hat\rho(\rv)=\int d\rv\sum_i\delta(\rv-\rv_i)=\sum_i 1=N$.  Hence one
obtains the (nontrivial) covariance of the value of $\hat A$ and the
total number $N$ of particles in the system, where we recall that the
latter is a fluctuating quantity in the grand ensemble that we use as
a foundation.

That the specific covariance form \eqref{EQchiAsCovariance} of
$\chi_A(\rv)$ is a physically meaningful measure of local fluctuations
of $\hat A(\rv^N)$ is suggested by several recent theoretical
developments that independently point to the relevance of this
specific type of correlation function. The arguably most prominent
example is the local compressibility $\chi_\mu(\rv)=\beta
\cov(\hat\rho(\rv), N)$, which is much advocated by Evans and
coworkers as a highly useful indicator for drying phenomena that occur
at a substrate \cite{evans2015jpcm, evans2019pnas, coe2022prl,
  coe2023}. In the present general framework we recover
$\chi_\mu(\rv)$ by the simple choice $\hat A=\beta N$ in
\eqr{EQchiAsCovariance}. The local compressibility $\chi_\mu(\rv)$
attains further significance as the parametric derivative
$\chi_\mu(\rv) = \partial\rho(\rv)/\partial \mu$ \cite{evans2015jpcm,
  evans2019pnas, coe2022prl, coe2023}, which hence measures the
changes in the equilibrium density profile upon changing the chemical
potential~(while keeping both the shape of the external potential and
temperature constant).

The motivation to treat the dependence on $T$ similarly to the
dependence on $\mu$ led Eckert {\it et al.}~\cite{eckert2020,
  eckert2023fluctuation} to correspondingly consider the thermal
susceptibility $\chi_T(\rv)=\partial \rho(\rv)/\partial T$, whereby
$\mu$ and again the shape of the external potential are kept fixed
upon differentiating. An equivalent covariance expression is
$\chi_T(\rv)=\cov(\hat \rho(\rv), \hat S)$, where the entropy operator
is $\hat S=-k_B \ln \Psi_\rmeq$ with the standard equilibrium
probability distribution $\Psi_\rmeq=\e^{-\beta (H-\mu N)}/\Xi$; see
Refs.~\cite{eckert2020, eckert2023fluctuation} for further
considerations that make $\chi_T(\rv)$ accessible in simulations.

While both of the above similarities are based on concrete choices for
$\hat A$, in the recent hyperforce theory by Robitschko {\it et
  al.}~\cite{robitschko2024any} the explicit form
\eqref{EQchiAsCovariance} of $\chi_A(\rv)$ features in the formulation
of exact Noether sum rules that emerge from thermal gauge invariance
\cite{mueller2024gauge}.  We lay out in detail the relationship of the
present hyperdensity functional approach to the hyperforce theory
below in Sec.~\ref{SEChyperforce}.

It is straightforward to show that the covariance
form~\eqref{EQchiAsCovariance} of the hyperfluctuation profile
$\chi_A(\rv)$ is generated in analogy to the mechanism used in
Refs.~\cite{evans2015jpcm, evans2019pnas, coe2022prl, eckert2020,
  coe2022pre, eckert2023fluctuation} from parametrically
differentiating the density profile \eqref{EQrhoAsAverage}. In the
present case we have
\begin{align}
  \chi_A(\rv) &= \frac{\partial \rho(\rv)}{\partial \lambda},
  \label{EQchiAsParametricDerivative}
\end{align}
where $\mu, T$ and the form of the external potential $V_\rmext(\rv)$
are all kept fixed upon building the derivative. That
\eqr{EQchiAsParametricDerivative} holds can be seen by explicit
calculation of the parameter derivative of the density profile. To do
so, we explicitly spell out the average \eqref{EQrhoAsAverage} to
obtain the density profile in the form $\rho(\rv)=\Tr \hat
\rho(\rv)\e^{-\beta(H-\mu N)+\lambda \hat A}/\Xi$.  The dependence on
$\lambda$ occurs both directly in the extended Boltzmann factor
$\e^{-\beta(H-\mu N)+\lambda\hat A}$ as well as in the extended
partition sum $\Xi=\Tr \e^{-\beta(H-\mu N)+\lambda\hat A}$. Upon
differentiating, the product rule gives two terms that constitute the
covariance~\eqref{EQchiAsCovariance}.  As before, we remain thereby
interested in the case $\lambda\to 0$ after having taken the
derivative in \eqr{EQchiAsParametricDerivative}. Hence the local
fluctuations of $\hat A$ are captured as they are generated from
interparticle coupling that the original Hamiltonian~$H$ generates in
the system.

\begin{figure}[htb]
  \includegraphics[width=0.45\textwidth]{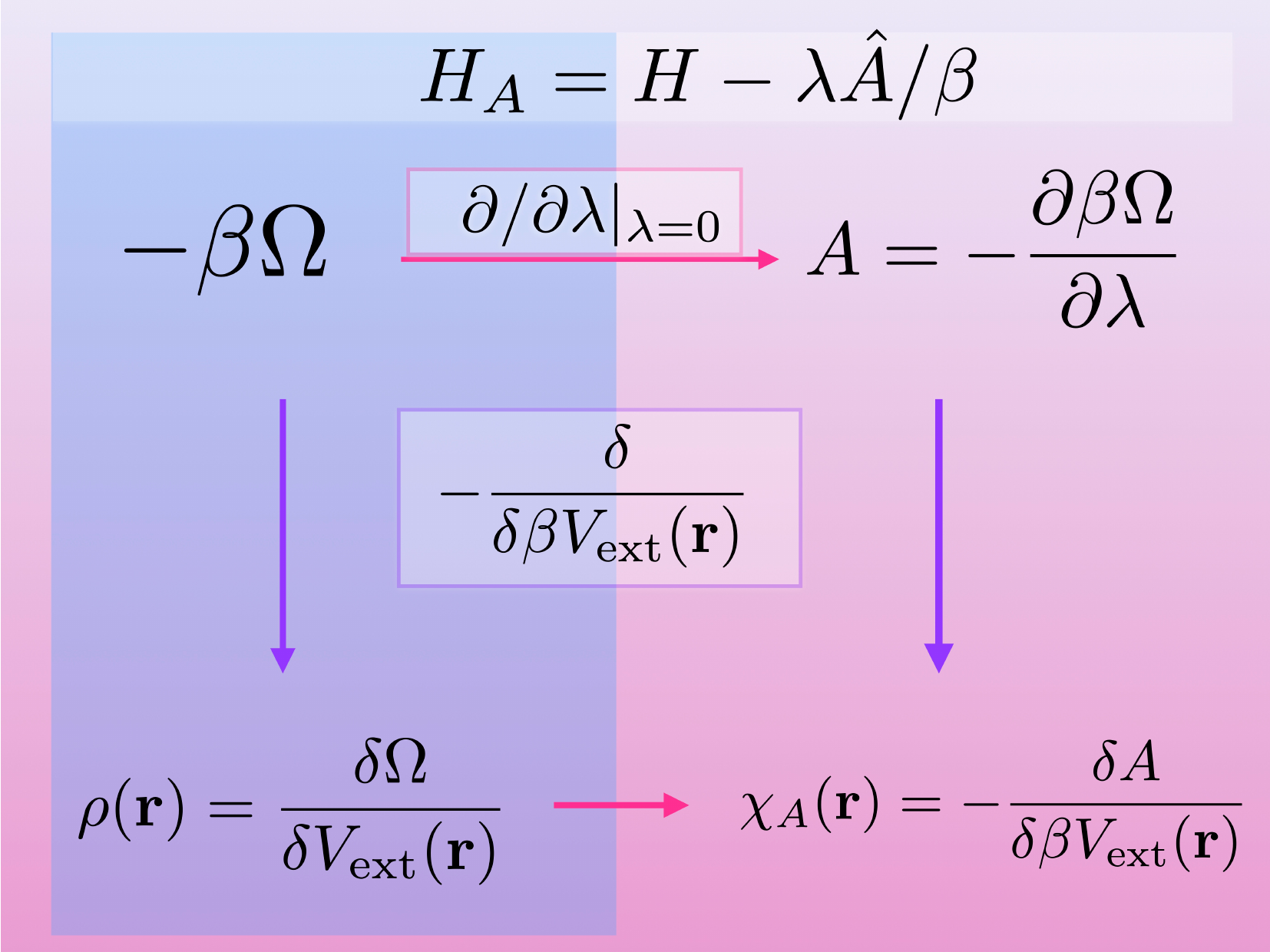}
    \caption{Illustration of parametric and functional relationships
      in the extended ensemble. The generalized Hamiltonian $H_A =
      H-\lambda \hat A/\beta$ is given by \eqr{EQhamiltonianA} and it
      has the associated scaled grand potential $-\beta\Omega$ and
      density profile $\rho(\rv)$. The mean $A$ is obtained via
      parametric differentiation according to $\partial/\partial
      \lambda|_{\lambda=0}$. Similarly the hyperfluctuation profile
      $\chi_A(\rv)$ follows according to
      \eqr{EQchiAsExternalPotentialDerivative}.
    \label{FIG2}
   }
\end{figure}

A further important mechanism to generate the hyperfluctuation profile
was identified by Eckert {\it et al.}~\cite{eckert2023fluctuation} in
the context of their investigation into the local compressibility and
thermal susceptibility. Upon investigating this specific case, they
have identified the following general mechanism:
\begin{align}
  \chi_A(\rv) = 
  -\frac{\delta A}{\delta \beta V_\rmext(\rv)}.
  \label{EQchiAsExternalPotentialDerivative}
\end{align}
Equation \eqref{EQchiAsExternalPotentialDerivative} is significant as
it reveals $\chi_A(\rv)$ as the response function of the average $A$
against changes in the (negative and thermally scaled) external
potential.

Proving \eqr{EQchiAsExternalPotentialDerivative} is straightforward
upon starting with $\chi_A(\rv)$ in the parametric derivative form
\eqref{EQchiAsParametricDerivative} and interchanging the order of
differentiation according to
\begin{align}
  \chi_A(\rv) &= \frac{\partial \rho(\rv)}{\partial \lambda} 
  = \frac{\partial}{\partial \lambda} 
  \frac{\delta \Omega}{\delta V_\rmext(\rv)}
  =\frac{\delta}{\delta  V_\rmext(\rv)}
  \frac{\partial \Omega}{\partial \lambda}
  \notag\\
  &= -\frac{\delta}{\delta \beta V_\rmext(\rv)}
  \Big(-\frac{\partial \beta \Omega}{\partial \lambda}\Big)
  = -\frac{\delta  A}{\delta \beta V_\rmext(\rv)}.
\end{align}
We have first re-written the density profile via the standard
functional response formula \cite{hansen2013, evans1979}
\begin{align}
  \rho(\rv)=\frac{\delta \Omega}{\delta V_\rmext(\rv)},
  \label{EQdeltaOmegaByDeltaVext}
\end{align}
and then identified the average $A=-\partial \beta \Omega/\partial
\lambda$ via \eqr{EQmeanAparametricDerivative}. Note that the standard
mechanism \eqref{EQdeltaOmegaByDeltaVext} to generate the density
profile $\rho(\rv)$ bears strong similarities with the generation
\eqref{EQchiAsExternalPotentialDerivative} of the hyperfluctuation
profile $\chi_A(\rv)$ via functional
differentiation. Figure~\ref{FIG2} depicts an illustration of the
mutual relationships.

\subsection{Mermin-Evans minimization principle in the extended ensemble}
\label{SECminimizationPrinciple}

We base the construction of the density functional formulation on the
extended Hamiltonian \eqref{EQhamiltonianA} and hence consider the
system as an interacting many-body system that is exposed to an
unchanged external potential $V_\rmext(\rv)$ as it occurs in
\eqr{EQhamiltonian}. Crucially, we attribute the differences between
the two Hamiltonian $H$ and $H_A$ solely to a change in interparticle
interactions. This implies going from the original interparticle
interaction potential $u(\rv^N)$ to the virtual interparticle
interaction potential $u_A(\rv^N)$ of the extended system, as it is
given by
\begin{align}
  u_A(\rv^N) &= u(\rv^N) - \lambda \hat A(\rv^N)/\beta.
  \label{EQuA}
\end{align}
Using \eqr{EQuA} to eliminate the explicit occurrence of $u(\rv^N)$ in
the extended Hamiltonian \eqref{EQhamiltonianA} then yields $H_A$ in
the standard form $H_A = \sum_i \pv_i^2/(2m)+ u_A(\rv^N) + \sum_i
V_\rmext(\rv_i)$. The same result is obtained from simply replacing
$u(\rv^N)$ by $u_A(\rv^N)$ in \eqr{EQhamiltonian}.

In typical systems $u(\rv^N)$ can be of very specific form, say being
composed of solely pair interaction contributions. Even if this is the
case, then the form of $u_A(\rv^N)$ can in general be much more
complex and in particular it can possess many-body contributions,
depending on the specific form of the dependence of $\hat
A(\rv^N)$. Furthermore additional one-body contributions, on top of
the bare external potential $V_\rmext(\rv)$, could be present. In
principle one can split off such terms and combine them with
$V_\rmext(\rv)$ into a modified external potential. We do not perform
this modification though and both for simplicity and conceptual
clarity keep $u_A(\rv^N)$ in its full form~\eqref{EQuA}. We hence
rather exploit that the general density functional framework poses no
restrictions on the form of the interparticle potential other than
that the resulting thermal ensemble needs to be well-defined. We
return to the cases of specific one- and two-body forms of the
observable $\hat A(\rv^N)$ in Sec.~\ref{SECclassifyingObservables}
after the development of the general framework.

Although we remain ultimately interested only in the limit $\lambda\to
0$, such that the extended Hamiltonian \eqref{EQhamiltonianA} reduces
to the Hamiltonian \eqref{EQhamiltonian} of the original system, we
start by formulating the grand potential density functional
$\Omega[\rho]$ for the generalized Hamiltonian~$H_A$. The general
variational principle \cite{mermin1965, evans1979, evans1992}
ascertains that
\begin{align}
  \frac{\delta \Omega[\rho]}{\delta\rho(\rv)} &= 0 \quad {\rm (min)},
  \label{EQomegaMinimal}
\end{align}
where equality holds at the minimum and the minimizer is the true
equilibrium density distribution $\rho(\rv)$ given as the thermal
average \eqref{EQrhoAsAverage}.  In our present formulation
$\rho(\rv)$ is the density profile of the extended system with
generalized interparticle potential $u_A(\rv^N)$ and in the presence
of the fixed external potential $V_\rmext(\rv)$ and at the fixed
thermodynamic statepoint $\mu, T$.

The standard form of the grand potential density functional
$\Omega[\rho]$ consists of a sum of ideal gas, excess (over ideal
gas), and external contributions according to \cite{evans1979,
  hansen2013}
\begin{align}
  \Omega[\rho] &= F_\rmid[\rho] + F_\rmexc[\rho]
  + \int d\rv \rho(\rv) [V_\rmext(\rv)-\mu].
  \label{EQomegaSplitting}
\end{align}
The intrinsic Helmholtz free energy functional of the ideal gas,
$F_\rmid[\rho]$, is exactly known in the form $F_\rmid[\rho]=k_BT \int
d\rv \rho(\rv)[\ln(\rho(\rv)\Lambda^d)-1]$, where $\Lambda$ denotes
the thermal de Broglie wavelength. The excess free energy functional
$F_\rmexc[\rho]$ is unkown in general and due to all remaining
interactions that are not accounted for by $V_\rmext(\rv)$. Hence in
the present setup $F_\rmexc[\rho]$ depends on and is generated by the
extended interparticle interaction potential $u_A(\rv^N)$, as is given
by \eqr{EQuA}. In particular, the dependence of $u_A(\rv^N)$ on the
coupling parameter $\lambda$ renders the excess free energy functional
$F_\rmexc[\rho]$ parametrically dependent on $\lambda$. In the limit
$\lambda\to 0$ the excess free energy functional of the original
system, with bare interparticle interaction potential $u(\rv^N)$, is
thereby restored by construction.

Inserting the grand potential density functional splitting
\eqref{EQomegaSplitting} into the minimization principle
\eqref{EQomegaMinimal} and calculating the functional derivative
yields the Euler-Lagrange equation of classical density functional
theory in the following standard form \cite{evans1979, hansen2013,
  schmidt2022rmp}:
\begin{align}
  c_1(\rv;[\rho]) &= \ln[\rho(\rv)\Lambda^d] 
  + \beta [V_\rmext(\rv) -  \mu].
  \label{EQel}
\end{align}
Thereby the one-body direct correlation functional $c_1(\rv;[\rho])$
is given as the density functional derivative of the scaled excess
free energy functional $-\beta F_\rmexc[\rho]$, such that
\begin{align}
  c_1(\rv;[\rho]) &= 
  -\frac{\delta \beta F_\rmexc[\rho]}{\delta\rho(\rv)}.
  \label{EQc1definition}
\end{align}
Clearly the ensemble generalization is imprinted into
$c_1(\rv;[\rho])$ through the generalized interparticle potential
$u_A(\rv^N)$ given via \eqr{EQuA}. As a result $c_1(\rv;[\rho])$
depends both on the value of the coupling parameter $\lambda$ as well
as on the specific form of the observable $\hat A(\rv^N)$.  Performing
the limit $\lambda\to 0$ restores the one-body direct correlation
functional of the original system, with bare interparticle interaction
potential $u(\rv^N)$.

Before we take the parametric limit though, we differentiate
\eqr{EQc1definition} with respect to $\lambda$ and hence define the
resulting hyperdirect correlation functional $c_A(\rv;[\rho])$ as
\begin{align}
  c_A(\rv;[\rho]) &= 
  \frac{\partial c_1(\rv;[\rho])}{\partial \lambda}\Big|_\rho,
  \label{EQcAasParametricDerivative}
\end{align}
where the density profile is kept fixed upon building the parametric
derivative, which hence acts exclusively on the implied Hamiltonian
$H_A$. In general $c_A(\rv;[\rho])$ will remain nonvanishing when the
limit $\lambda\to 0$ is performed in \eqr{EQcAasParametricDerivative}
after the parametric derivative is taken. This procedure both restores
the statistical mechanics of the original system, but as we will
demonstrate in the following, it also allows one to capture the
relevant statistical information of the behaviour of $\hat A(\rv^N)$
that the original system displays via the one-body hyperdirect
correlation functional $c_A(\rv;[\rho])$.

The functional derivative relationship~\eqref{EQc1definition} can be
inverted straightforwardly by functional integration. The general
functional integral $-\beta F_\rmexc[\rho]=\int {\cal
  D}[\rho]c_1(\rv;[\rho])$ can be parameterized efficiently according
to \cite{evans1979, evans1992, sammueller2023whyNeural}:
\begin{align}
  -\beta F_\rmexc[\rho] &=
  \int d\rv \rho(\rv) \int_0^1 da c_1(\rv;[a\rho]).
  \label{EQFexcViaFunctionalIntegral}
\end{align}
The functional argument $a\rho(\rv)$ of the one-body direct
correlation functional is thereby merely a scaled version of the
``target'' density profile $\rho(\rv)$ \cite{evans1979,
  evans1992}. Equation \eqref{EQFexcViaFunctionalIntegral} is suitable
for numerical evaluation upon providing a specific form of
$c_1(\rv;[\rho])$ for the system under consideration
\cite{sammueller2023neural, sammueller2023whyNeural,
  sammueller2023neuralTutorial, sammueller2024hyperDFT,
  sammueller2024attraction, sammueller2024pairmatching}.  Analogously,
functional integration is relevant for evaluating the density
functional $A[\rho]$ given a hyperdirect correlation functional
$c_A(\rv;[\rho])$, as described in more detail in
Sec.~\ref{SEChyperDensityFunctionals}.

\subsection{Levy's constrained search in the extended ensemble}
\label{SEClevy}

The existence and uniqueness of the grand potential density functional
\eqref{EQomegaSplitting} is typically proven by contradiction
\cite{evans1979, hansen2013}.  Levy's constrained search method
\cite{levy1979, kohn1999nobel} provides an arguably more constructive
route and the method applies classically as well \cite{dwandaru2011,
  schmidt2022rmp}. Briefly, one starts from the standard Mermin-Evans
many-body functional \cite{mermin1965, evans1979} $\Omega_M[\Psi] =
\Tr \Psi (H_A-\mu N + k_BT\ln \Psi)$ where $\Psi(\rv^N,\pv^N)$ is a
many-body distribution function which is normalized according to $\Tr
\Psi=1$ but is otherwise general.  Splitting off from $\Omega_M[\Psi]$
the contributions from chemical potential and external potential
leaves over an intrinsic many-body functional $F_M[\Psi]$ given by
\begin{align}
  F_M[\Psi] &= 
  \Tr \Big(\sum_i\frac{\pv_i^2}{2m} 
  + u_A(\rv^N) + k_BT \ln\Psi \Big) \Psi.
  \label{EQFMerminEvans}
\end{align}
The Levy method generates from \eqr{EQFMerminEvans} the corresponding
free energy density functional $F[\rho]=F_\rmid[\rho]+F_\rmexc[\rho]$
via the following constrained search:
\begin{align}
  F[\rho] &= \min_{\Psi\to\rho} F_M[\Psi].
  \label{EQFexcViaLevy}
\end{align}
Here $\Psi(\rv^N,\pv^N)$ is constrained to generate the prescribed
``target'' density profile $\rho(\rv)$, which is the functional
argument on the left hand side of \eqr{EQFexcViaLevy}. The constraint,
indicated as $\Psi\to\rho$ in \eqr{EQFexcViaLevy}, enforces the
following identity
\begin{align}
  \rho(\rv) &= \Tr \hat\rho(\rv) \Psi,
\end{align}
where the density operator $\hat\rho(\rv)$ is defined in
\eqr{EQdensityOperator}.

It is interesting to note that the mean of the considered observable,
for given form of $\Psi(\rv^N,\pv^N)$ is obtained as the following
partial derivative:
\begin{align}
  A = -\frac{\partial F_M[\Psi]}{\partial \lambda}\Big|_{\Psi,\mu,T},
\end{align}
where the value $A$ is the intended one provided that
$\Psi(\rv^N,\pv^N)$ has the correct equilibrium form. That no further
terms become relevant when parametrically differentiating is not
immediately obvious and we postpone the construction of the density
functional $A[\rho]$ to Sec.~\ref{SEChyperDensityFunctionals} below.

\subsection{Hyper-Ornstein-Zernike relation}
\label{SEChyperOZgeneral}

While we have introduced microscopic expressions for the
hyperfluctuation profile~$\chi_A(\rv)$ both via the covariance
\eqref{EQchiAsCovariance} and, equivalently, by the parametric
derivative~\eqref{EQchiAsParametricDerivative}, this particular
correlation function emerges also very naturally within the present
hyperdensity functional treatment based on a hyper-Ornstein-Zernike
relation, as we will demonstrate in the following. We recall that the
two-body Ornstein-Zernike relationship is a staple of liquid state
theory and that its origin lies, together with the introduction of
pair direct correlation functions, in the treatment of critical
opalescence by Ornstein and Zernike in 1914 \cite{hansen2013,
  ornstein1914}.

The two-body direct correlation functional is given as the following
second density functional derivative:
\begin{align}
  c_2(\rv,\rv';[\rho])
  &=-\frac{\delta^2 \beta F_\rmexc[\rho]}
  {\delta\rho(\rv)\delta\rho(\rv')}.
  \label{EQc2fromFexc}
\end{align}
Expressing one of the two chained functional derivatives in
\eqr{EQc2fromFexc} via \eqr{EQc1definition} leads to the following
alternative and equivalent form:
\begin{align}
  c_2(\rv,\rv';[\rho])&=
  \frac{\delta c_1(\rv;[\rho])}{\delta\rho(\rv')}.
  \label{EQc2fromc1}
\end{align}

A complementary hierarchy of total correlation functions is obtained
from functionally differentiating the elementary grand potential with
respect to $V_\rmext(\rv)$ \cite{evans1979, hansen2013}, such that
$\delta^2\Omega/[\delta V_\rmext(\rv)\delta V_\rmext(\rv')] =
\beta\cov(\hat\rho(\rv),\hat\rho(\rv'))=\beta H_2(\rv,\rv')$, where
$H_2(\rv,\rv')$ is the standard two-body correlation function of
density fluctuations \cite{hansen2013, evans1979}. The inhomogeneous
Ornstein-Zernike equation then relates the total and direct
correlation functions on the two-body level according to
\cite{evans1979, hansen2013, schmidt2022rmp}:
\begin{align}
  H_2(\rv,\rv')
  &= \rho(\rv)\delta(\rv-\rv')
  \notag\\ &\quad
  + \rho(\rv)\int d\rv'' c_2(\rv,\rv'';[\rho]) H_2(\rv'',\rv').
  \label{EQozStandardTwoBody}
\end{align}

To derive the hyper-Ornstein-Zernike relation, we follow a formal and
very direct strategy based on the Euler-Lagrange equation
\cite{schmidt2022rmp, brader2013noz, brader2014noz, eckert2020,
  eckert2023fluctuation}. The method is based on exploiting the
generality of the minimization condition \eqref{EQel} and the concept
that this remains a valid identity upon suitable
differentiation. Carrying out functional derivatives thereby requires
to take account of both direct changes as well as mediated changes
that are generated via changes in the density profile itself. We refer
the Reader to Ref.~\cite{schmidt2022rmp} for the corresponding
derivation of the two-body Ornstein-Zernike relation
\eqref{EQozStandardTwoBody} from functionally differentiating the
standard Euler-Lagrange equation \eqref{EQel} with respect
to~$V_\rmext(\rv')$.

Here we turn to the extended ensemble and differentiate the
Euler-Lagrange equation~\eqref{EQel} with respect to the coupling
parameter~$\lambda$. We consider both the one-body direct correlation
functional $c_1(\rv;[\rho])$ and the equilibrium density profile
$\rho(\rv)$ to be those corresponding to the extended
Hamiltonian~$H_A$.  The external potential $V_\rmext(\rv)$ and the
statepoint $\mu, T$ remain prescribed. Hence the Euler-Lagrange
equation \eqref{EQel} is satisfied for a range of values of
$\lambda$. We hence retain a valid identity upon differentiating
\eqr{EQel} with respect to $\lambda$. Respecting the involved
dependencies yields the following hyper-Ornstein-Zernike equation:
\begin{align}
  c_A(\rv;[\rho]) &= \frac{\chi_A(\rv)}{\rho(\rv)}
  -\int d\rv' c_2(\rv,\rv';[\rho]) \chi_A(\rv'),
  \label{EQhyperOZ}
\end{align}
where the one-body hyperdirect correlation functional
$c_A(\rv;[\rho])$ remains being defined via the parametric derivative
\eqref{EQcAasParametricDerivative}. 

We provide a detailed derivation of \eqr{EQhyperOZ} in the
following. We first consider the left hand side of the Euler-Lagrange
equation \eqref{EQel}. Differentiating the occurring one-body direct
correlation functional $c_1(\rv;[\rho])$ with respect to $\lambda$
yields
\begin{align}
  \frac{\partial c_1(\rv;[\rho])}{\partial \lambda}
  \Big|_{V_\rmext}&=
  \frac{\partial c_1(\rv;[\rho])}{\partial \lambda}\Big|_{\rho}
  \notag\\&\quad
  + \int d\rv' \frac{\delta c_1(\rv;[\rho])}{\delta\rho(\rv')}
  \frac{\partial\rho(\rv')}{\partial\lambda}  \Big|_{V_\rmext}
  \label{EQhyperOZderivation1}\\
  &= c_A(\rv;[\rho]) + \int d\rv' c_2(\rv,\rv';[\rho]) \chi_A(\rv')
  \label{EQhyperOZderivation2}.
\end{align}  
The two terms in \eqr{EQhyperOZderivation1} arise from the direct
changes upon changing $\lambda$ (first term) and from the changes that
are mediated by alteration of the density profile and using the
functional chain rule (second term). To obtain
\eqr{EQhyperOZderivation2} we have identified the two parametric
derivatives as the one-body hyperdirect correlation functional
$c_A(\rv;[\rho])$ via \eqr{EQcAasParametricDerivative} and the
hyperfluctuation profile $\chi_A(\rv)$ via
\eqr{EQchiAsParametricDerivative}.

It remains to also differentiate the right hand side of the
Euler-Lagrange equation \eqref{EQel} with respect to $\lambda$. We
recall that the statepoint $\mu, T$ and the shape of the external
potential $V_\rmext(\rv)$ are kept fixed upon changing the parameter
$\lambda$. Hence the corresponding terms vanish upon differentiating
and the result is compact:
\begin{align}
  \frac{\partial \ln[\rho(\rv)\Lambda^d]}{\partial \lambda}
   + \frac{\partial [\beta V_\rmext(\rv)-\beta\mu]}{\partial \lambda}
   &= \frac{1}{\rho(\rv)}
   \frac{\partial \rho(\rv)}{\partial \lambda}
   =\frac{\chi_A(\rv)}{\rho(\rv)},
   \label{EQhyperOZderivation3}
\end{align}  
where in the last step we have again identified $\chi_A(\rv)$ via
\eqr{EQchiAsParametricDerivative}. The exact sum rule then follows
from equating the right hand sides of
Eqs.~\eqref{EQhyperOZderivation2} and \eqref{EQhyperOZderivation3},
which yields the hyper-Ornstein-Zernike equation in the form
\eqref{EQhyperOZ} after re-arranging.

The hyper-Ornstein-Zernike equation \eqref{EQhyperOZ} is exact and in
a similar fashion as the inhomogeneous two-body Ornstein-Zernike
equation of liquid state theory \cite{hansen2013} it can assume a
multitude of different roles. That \eqr{EQhyperOZ} only depends on a
single free position argument, rather than two space points, as are
occurring in \eqr{EQozStandardTwoBody}, is in keeping with the
fluctuation Ornstein-Zernike relations for $\chi_\mu(\rv)$ and for
$\chi_T(\rv)$ \cite{eckert2020, eckert2023fluctuation}. Furthermore,
similarly to \eqr{EQhyperOZ} these exact relationships also feature
the mediation of fluctuations via spatial integration of the standard
two-body direct correlation functional $c_2(\rv,\rv';[\rho])$; we
refer the Reader to Ref.~\cite{eckert2023fluctuation} for further
details.

In the following we describe two application scenarios of the
hyper-Ornstein-Zernike relation \eqref{EQhyperOZ} with a specific
choice of the observable $\hat A(\rv^N)$ being under consideration. We
assume that the density profile~$\rho(\rv)$ for the bare Hamiltonian
$H$ has been obtained from either solution of the Euler-Lagrange
equation \eqref{EQel} or from the equilibrium average
\eqref{EQrhoAsAverage} as carried out, e.g., in grand canonical Monte
Carlo simulations. In both cases the form of $V_\rmext(\rv)$ has been
given and the thermodynamic statepoint~$\mu, T$ has been
prescribed. The situation is hence similar to application of the
standard inhomogeneous two-body Ornstein-Zernike equation
\eqref{EQozStandardTwoBody} where knowledge of the density profile is
typically also required. Furthermore we assume that the two-body
direct correlation functional $c_2(\rv,\rv';[\rho])$ is
available. This can come either from the standard second functional
derivative \eqref{EQc2fromFexc} of a (typically approximate)
analytical model form of the excess free energy functional
$F_\rmexc[\rho]$. Alternatively, the neural functional calculus of
Refs.~\cite{sammueller2023neural, sammueller2023whyNeural,
  sammueller2023neuralTutorial} provides direct representation of
$c_1(\rv;[\rho])$ as a trained neural network and performing automatic
differentiation~\cite{baydin2018autodiff, sammueller2023neural,
  sammueller2023whyNeural, sammueller2023neuralTutorial} then creates
the neural two-body direct correlation functional
$c_2(\rv,\rv';[\rho])$ via the first functional derivative
\eqref{EQc2fromc1}.

We first assume that both the hyperdirect correlation functional
$c_A(\rv;[\rho])$ and the two-body direct correlation functional
$c_2(\rv,\rv';[\rho])$ are known.  The particles interact thereby
solely via the bare interaction potential $u(\rv^N)$ and hence
$\lambda\to 0$ has been taken. Then evaluating $c_A(\rv;[\rho])$ at
the equilibrium density $\rho(\rv)$ turns the left hand side of the
hyper-Ornstein-Zernike equation~\eqref{EQhyperOZ} into a fixed spatial
inhomogeneity $c_A(\rv)$. The remaining terms on the right hand side
need to accommodate this inhomogeneity and hence \eqr{EQhyperOZ}
constitutes an integral equation for the hyperfluctuation profile
$\chi_A(\rv)$. Here $c_2(\rv,\rv';[\rho])$, when evaluated at the
equilibrium density profile, yields a generalized, i.e., fully
position-dependent, convolution kernel $c_2(\rv,\rv')$. Besides the
(standard) input of the density profile and $c_2(\rv,\rv';[\rho])$,
this requires mere availability of the hyperdirect correlation
functional $c_A(\rv;[\rho])$, which points to the prowess of this
object and we recall its definition via the parametric derivative
\eqref{EQcAasParametricDerivative} of the extended one-body direct
correlation functional. See Ref.~\cite{sammueller2024attraction} for
the application of this route to the calculation of the local
compressibility.

The second use of the hyper-Ornstein-Zernike
relation~\eqref{EQhyperOZ} targets the construction of
$c_A(\rv;[\rho])$. The method is based on simulations, where the
hyperfluctuation profile $\chi_A(\rv)$ is available via sampling the
covariance~\eqref{EQchiAsCovariance}. We assume that this has been
accomplished together with sampling of $\rho(\rv)$, again for given
$V_\rmext(\rv), \mu, T$. We also assume that the two-body direct
correlation functional $c_2(\rv,\rv';[\rho])$ is known by one of the
two methods described above (or alternatively by inhomogeneous liquid
integral equation theory). Then all terms on the right hand side of
\eqr{EQhyperOZ} can be evaluated for the specific system under
consideration, as specified by its Hamiltonian $H$, including the form
of the external potential, and the thermodynamic parameters
$\mu,T$. For the specific chosen observable $\hat A(\rv^N)$ the right
hand side of \eqr{EQhyperOZ} can hence be evaluated and the one-body
hyperdirect correlation function $c_A(\rv)$ for the specific system
under consideration has become available. While knowing the function
in a specific case does not yet imply knowledge of the functional
$c_A(\rv;[\rho])$, creating a set of such profiles from simulating the
system under a range of different conditions is the basis for using
supervised machine learning following
Refs.~\cite{sammueller2023neural, sammueller2023whyNeural,
  sammueller2023neuralTutorial, sammueller2024pairmatching,
  sammueller2024attraction} in the construction of neural hyperdirect
correlation functionals $c_A(\rv;[\rho])$
\cite{sammueller2024hyperDFT}. Details and concrete applications of
this technique are given in Sec.~\ref{SECapplications}.

Having demonstrated the intimate links between the
hyper-Ornstein-Zernike equation \eqref{EQhyperOZ}, the hyperdirect
correlation functional \eqref{EQcAasParametricDerivative} and the
hyperfluctuation profile \eqref{EQchiAsCovariance}, we next turn to
addressing the mean $A$ of the chosen observable, given via the
elementary form \eqref{EQmeanAasAverage}, in the density functional
context.

\subsection{Any observable as a hyperdensity functional}
\label{SEChyperDensityFunctionals}

We aim at a density functional representation of the average $A$. We
start from the parametric differentiation according to
\eqr{EQmeanAparametricDerivative}.  However, rather than using the
elementary form of the grand potential, $\Omega=-k_BT\ln\Xi$, we work
on the basis of the grand potential functional $\Omega[\rho]$. The
splitting of $\Omega[\rho]$ into its additive constituents via
\eqr{EQomegaSplitting} enables one to carry out the required
parametric derivative with respect to $\lambda$ under identical
conditions as in \eqr{EQmeanAparametricDerivative}, i.e., at fixed
form of $V_\rmext(\rv)$ and fixed statepoint $\mu, T$. The result
(derived below) is compact:
\begin{align}
  A[\rho] = -\frac{\partial \beta F_\rmexc[\rho]}{\partial \lambda}
  \Big|_\rho,
  \label{EQmeanAasFunctional}
\end{align}
where $F_\rmexc[\rho]$ is the excess free energy functional generated
by the extended Hamiltonian $H_A$ and the density profile $\rho(\rv)$
is kept fixed upon differentiating. As before we can take the limit
$\lambda\to 0$ after the derivative is taken, in order to restore the
statistical mechanics of the original Hamiltonian $H$. No other terms
in \eqr{EQomegaSplitting} depend explicitly on~$\lambda$, which leads
to the simplicity of the right hand side of \eqr{EQmeanAasFunctional}.

As \eqr{EQmeanAasFunctional} is a central and arguably
counter-intuitive result, we provide more elementary details for its
derivation. Taking into account the definition of $A$ via
\eqr{EQmeanAparametricDerivative}, we start from the grand potential
functional $\Omega[\rho]$, as expressed via the splitting
\eqref{EQomegaSplitting}, and differentiate with respect to the
coupling parameter $\lambda$ as follows:
\begin{align}
  \frac{\partial \Omega[\rho]}{\partial \lambda}  \Big|_{V_\rmext} &=
  \frac{\partial \Omega[\rho]}{\partial \lambda}\Big|_{\rho}
  + \int d\rv \frac{\delta \Omega[\rho]}{\delta\rho(\rv)}
  \Big|_{V_\rmext}
  \chi_A(\rv).
  \label{EQmeanAderivation1}
\end{align}
The first term on the right hand side of \eqr{EQmeanAderivation1}
accounts for the ``inherent'' changes in $\Omega[\rho]$ that occur
upon keeping the density profile fixed, as is indicated in the
notation. This term can be pictured as $\lambda$ controlling the
extended Hamiltonian $H_A$, see \eqr{EQhamiltonianA}, upon keeping the
density profile fixed when changing $H_A$. By splitting $\Omega[\rho]$
according to \eqr{EQomegaSplitting} we can simplify as follows:
\begin{align}
  \frac{\partial\Omega[\rho]}{\partial \lambda} \Big|_{\rho} &=
  \frac{\partial F_\rmexc[\rho]}{\partial \lambda} \Big|_{\rho}
  \notag \\ & \qquad +
  \frac{\partial}
       {\partial \lambda} \Big|_{\rho}
       \Big(F_\rmid[\rho] +  \int d\rv \rho(\rv)[V_\rmext(\rv)-\mu]
       \Big)  \label{EQmeanAderivation2}\\       
  &= \frac{\partial F_\rmexc[\rho]}{\partial \lambda} \Big|_{\rho},
  \label{EQmeanAderivation3}
\end{align}
where the ideal, external, and chemical contributions in
\eqr{EQmeanAderivation2} carry no dependence on $\lambda$ and hence
vanish under the parametric derivative, which leads to
\eqr{EQmeanAderivation3}.

The second contribution on the right hand side of
\eqr{EQmeanAderivation1} arises from the chain rule and identifying
the hyperfluctuation profile $\chi_A(\rv)$ via the definition
\eqref{EQchiAsParametricDerivative} as the partial derivative
$\partial\rho(\rv)/\partial \lambda$. However, it follows
straightforwardly that
\begin{align}
  \int d\rv \frac{\delta \Omega[\rho]}{\delta\rho(\rv)} \Big|_{V_\rmext}
  \chi_A(\rv)
  &= 0.
  \label{EQmeanAderivation4}
\end{align}
This simplification is due to the first term inside of the integrand
vanishing identically: $\delta
\Omega[\rho]/\delta\rho(\rv)|_{V_\rmext}=0$, where, as indicated, the
derivative is taken at fixed $V_\rmext(\rv)$ and at constant value of
the coupling parameter $\lambda$ and fixed statepoint $\mu, T$. Due to
the fundamental variational principle~\eqref{EQomegaMinimal} in the
extended ensemble the result vanishes identically. Collecting the
results \eqref{EQmeanAderivation3} and \eqref{EQmeanAderivation4}
leaves over only the compact right hand side of
\eqr{EQmeanAasFunctional}.

Certainly the simplicity of \eqr{EQmeanAasFunctional} points towards
re-affirming the central role of the excess free energy functional
within the density functional framework as encapsulating the essence
of the interparticle coupling. The extended ensemble thereby
facilitates to both retain this encapsulation, but also to extend
towards a general observable $\hat A$.

We can now establish a connection between $A[\rho]$ and
$c_A(\rv;[\rho])$. Inserting \eqr{EQc1definition} into
\eqr{EQcAasParametricDerivative}, exchanging the orders of the
functional density derivative and the parametric derivative with
respect to $\lambda$, and identifying the density functional $A[\rho]$
via \eqr{EQmeanAasFunctional} yields:
\begin{align}
  c_A(\rv;[\rho]) &= \frac{\delta A[\rho]}{\delta\rho(\rv)}.
  \label{EQcAasFunctionalDerivative}
\end{align}
Equation \eqref{EQcAasFunctionalDerivative} adds further significance
to the hyperdirect correlation functional as reflecting the changes in
the mean $A$ upon changing the density profile. As laid out and
explicitly used in the derivation, the changes in the density profile
are monitored in the system at fixed thermodynamic conditions and the
particles interact solely via the original interaction potential
$u(\rv^N)$.

As we have alluded to above, the practical access to $c_A(\rv;[\rho])$
via supervised machine learning is arguably more direct than
attempting to construct $A[\rho]$ from first principles. Once the
one-body hyperdirect correlation functional is known, one can
straightforwardly carry out a functional integral, which we write
first in a formal way in the form
\begin{align}
  A[\rho] = \int {\cal D}[\rho]c_A(\rv;[\rho]).
  \label{EQmeanAviaFunctionalIntegralAbstract}
\end{align}
The functional integral operator $\int {\cal D}[\rho]$ performs a line
integral in the space of density functions; we refer the Reader to the
classical presentations by Evans \cite{evans1979, evans1992} and to
Ref.~\cite{sammueller2023whyNeural} for the implications in the light
of the neural functional theory. In practice carrying out the formal
functional integral \eqref{EQmeanAviaFunctionalIntegralAbstract}
requires to choose a parametrizaion, e.g., according to a simple
scaling:
\begin{align}
  A[\rho] &= \int d\rv \rho(\rv) \int_0^1 da c_A(\rv;[a\rho]).
  \label{EQmeanAviaFunctionalIntegral}
\end{align}
Here the functional argument $a\rho(\rv)$, with parameter $0\leq a
\leq 1$, of the hyperdirect correlation functional is a scaled version
of the ``target'' density profile $\rho(\rv)$ that is the argument on
the left hand side of \eqr{EQmeanAviaFunctionalIntegral}.  Equation
\eqref{EQmeanAviaFunctionalIntegral} thereby mirrors closely the
functional integration of $c_1(\rv;[\rho])$ to obtain the excess free
energy via \eqr{EQFexcViaFunctionalIntegral}.  When working with
neural functionals, evaluating \eqr{EQmeanAviaFunctionalIntegral}
numerically is a very fast operation, similar in performance to the
corresponding functional integral~\eqref{EQFexcViaFunctionalIntegral}
over $c_1(\rv;[\rho])$ \cite{sammueller2023neural,
  sammueller2023whyNeural, sammueller2023neuralTutorial,
  sammueller2024pairmatching, sammueller2024attraction,
  bui2024neuralrpm, kampa2024meta}.

\subsection{Wall hypercontact theorem}
\label{SECcontactTheorem}
As a specific situation, we consider a semi-infinite system, where a
hard wall at $x=0$ constrains all particle coordinates $\rv_i$ such
that all $x_i>0$. For large distances from the wall, $x\to \infty$,
the system approaches bulk with vanishing external potential. The bulk
pressure $p$ then characterizes the system and it is related to the
density $\rho(0^+)$ at the wall by the contact theorem
\cite{hansen2013, tschopp2022forceDFT},
\begin{align}
  \rho(0^+) &= \beta p.
  \label{EQhardWallContactDensityTheorem}
\end{align}
In generalization of the procedure by Evans and Stewart of changing
$\mu$~\cite{evans2015jpcm}, as followed up by Eckert {\it et al.}\ who
considered changing $T$~\cite{eckert2023fluctuation}, we here
parametrically differentiate the hard wall contact theorem
\eqref{EQhardWallContactDensityTheorem} with respect to the coupling
parameter~$\lambda$. The result is
\begin{align}
  \chi_A(0^+) &= \frac{A_b}{V},
  \label{EQhardWallContactChiTheorem}
\end{align}
where $\chi_A(0^+)$ is the contact value of the fluctuation profile
$\chi_A(\rv)$, the bulk expectation value of $\hat A(\rv^N)$ is
indicated by $A_b$, and $V$ is the bulk volume. The left hand side of
the contact theorem \eqref{EQhardWallContactChiTheorem} is obtained
from using the parametric derivative form
\eqref{EQchiAsParametricDerivative} of the hyperfluctuation
profile. The right hand side of \eqr{EQhardWallContactChiTheorem}
follows from noting that in bulk $\Omega=-pV$ and expressing the mean
$A$ via the parametric derivative \eqref{EQmeanAparametricDerivative}
of the grand potential, which gives $A_b = -\partial \beta \Omega /
\partial \lambda = V \partial \beta p /\partial \lambda = V \partial
\rho(0^+)/\partial \lambda = V \chi_A(0^+)$.

This concludes our presentation of the essentials of the hyperdensity
functional theory of Ref.~\cite{sammueller2024hyperDFT}. We proceed to
demonstrating the close relationship with the hyperforce framework by
Robitschko {\it et al.}~\cite{robitschko2024any}. This theory has
rather different roots, as it arises from the application of Noether's
theorem \cite{hermann2021noether, hermann2022topicalReview,
  hermann2022variance, hermann2022quantum, tschopp2022forceDFT,
  sammueller2023whatIsLiquid, hermann2023whatIsLiquid,
  robitschko2024any, mueller2024gauge} to the gauge invariance of
statistical mechanics \cite{mueller2024gauge}.  Similarly to the
present hyperdensity functional theory, however, this approach also
enables one to independently choose the Hamiltonian $H$ and the
observable~$\hat A$ under investigation.

\section{Hyperforce correlations}
\label{SEChyperforce}

\subsection{Local hyperforce balance}
\label{SEClocalHyperForceBalance}

The hyperdensity functional theory described in Sec.~\ref{SEChyperDFT}
is based on the standard density functional theory concepts of
representing the correlated many-body physics in terms of generating
functionals. Examples of their use are the generation of the direct
correlation functionals from the excess free energy functional
$F_\rmexc[\rho]$ via \eqr{EQc1definition} and the locally resolved
equilibrium balance condition in the form of the Euler-Lagrange
equation \eqref{EQel} that expresses the spatial homogeneity of the
sum of all contributions to the chemical potential. Although the
underlying Hamiltonian generates forces and the classical mechanics
crucially rests on the concept of forces, typically these do not
feature prominently in accounts of density functional theory, although
there are notable exceptions \cite{schmidt2022rmp,
  tschopp2022forceDFT, sammueller2022forceDFT}.

Here we aim to introduce forces into the hyperdensity functional
framework and thereby work on the basis of the extended Hamiltonian
$H_A$ together with the standard grand canonical setting at chemical
potential $\mu$ and temperature $T$, as described in
Sec.~\ref{SECextendedEnsemble}. The extended Hamiltonian~$H_A$
contains the extended interparticle interaction potential given by
\eqr{EQuA}, which we reproduce as $u_A(\rv^N) = u(\rv^N) - \lambda
\hat A(\rv^N)/\beta$. The corresponding extended one-body
interparticle force density operator is then defined as
\begin{align}
  \hat \Fv_{\rm int}(\rv) =  -\sum_i \delta(\rv-\rv_i) 
  \nabla_i u_A(\rv^N),
  \label{EQforceDensityOperator}
\end{align}
where the spatial localization is provided by the Dirac distribution,
similarly to the mechanism in the density operator
\eqref{EQdensityOperator}.  The mean equilibrium force density balance
then attains the standard form \cite{schmidt2022rmp}:
\begin{align}
  \beta \Fv_\rmint(\rv) 
  &= \nabla\rho(\rv)  + \rho(\rv) \nabla  \beta V_\rmext(\rv).
  \label{EQforceDensityBalance}
\end{align}
The mean one-body interparticle force density distribution is the
average
\begin{align}
  \Fv_\rmint(\rv)=\langle \hat \Fv_\rmint(\rv)\rangle,
  \label{EQFintAsMean}
\end{align}
with $\hat \Fv_\rmint(\rv)$ given by \eqr{EQforceDensityOperator} and
the density profile $\rho(\rv)=\langle \hat\rho(\rv) \rangle$
according to \eqr{EQrhoAsAverage}, with all averages being those of
the extended ensemble.

By construction the extended force density
balance~\eqref{EQforceDensityBalance} holds for any value of
$\lambda$. We proceed very similarly to our above strategy in
Sec.~\ref{SEChyperOZgeneral}, where we differentiated the
Euler-Lagrange equation \eqref{EQel} for the extended system with
respect to $\lambda$ to derive the hyper-Ornstein-Zernike relation
\eqref{EQhyperOZ}.

Here we parametrically differentiate the force density
balance~\eqref{EQforceDensityBalance} with respect to $\lambda$, which
yields a nontrivial identity as we will demonstrate. We proceed
stepwise and first address the interparticle force density
operator~\eqref{EQforceDensityOperator}. Taking account of the
form~\eqref{EQuA} of $u_A(\rv^N)$ yields the result
\begin{align}
  \frac{\partial \beta\hat\Fv_\rmint(\rv)}{\partial\lambda} 
  &= \hat\Sv_A(\rv),
  \label{EQgradA}
\end{align}
where we follow M\"uller {\it et al.}~\cite{mueller2024gauge,
  mueller2024whygauge} in defining the hyperforce density ``operator''
(phase space function) as
\begin{align}
  \hat\Sv_A(\rv) = \sum_i \delta(\rv-\rv_i) \nabla_i \hat A(\rv^N).
  \label{EQhatSAdefinition}
\end{align}
We next differentiate the entire average on the left hand side of
\eqr{EQforceDensityBalance}, which gives via the product rule:
\begin{align}
  \frac{\partial \beta \Fv_\rmint(\rv)}{\partial \lambda}
  &= \cov(\beta\hat \Fv_\rmint(\rv),\hat\rho(\rv))
  + \Sv_A(\rv),
  \label{EQFintParametricDerivative}
\end{align}
where the mean hyperforce density is $\Sv_A(\rv)=\langle \hat
\Sv_A(\rv)\rangle$ with $\hat \Sv_A(\rv)$ given by
\eqr{EQhatSAdefinition}.  The covariance on the right hand side of
\eqr{EQFintParametricDerivative} arises from differentiating the mean
$\langle \hat \Fv_\rmint(\rv) \rangle$ upon keeping the extended force
density operator \eqref{EQforceDensityOperator} itself fixed. The
second term on the right hand side of
\eqr{EQFintParametricDerivative}, i.e., the hyperforce density
$\Sv_A(\rv)$, stems from differentiating the operator
$\hat\Fv_\rmint(\rv)$, according to \eqr{EQgradA}, inside of the
average.

It remains to also parametrically differentiate the two terms on the
right hand side of \eqr{EQforceDensityBalance}, which respectively
gives
\begin{align}
  \frac{\partial}{\partial\lambda} \nabla\rho(\rv) &= \nabla\chi_A(\rv),
  \label{EQpartialLambda1}\\
  \frac{\partial}{\partial \lambda}\rho(\rv)\nabla\beta V_\rmext(\rv)
  &=  \chi_A(\rv)\nabla\beta V_\rmext(\rv).
  \label{EQpartialLambda2}
\end{align}
Carrying out the derivatives on the left hand sides of
Eqs.~\eqref{EQpartialLambda1} and \eqref{EQpartialLambda2} is
performed as before with both $\beta$ and $V_\rmext(\rv)$ being kept
fixed.  In both cases the parametric derivative of $\rho(\rv)$ with
respect to $\lambda$ then generates the hyperforce fluctuation profile
$\chi_A(\rv)$ via \eqr{EQchiAsParametricDerivative}.

In summary, using the results \eqref{EQFintParametricDerivative},
\eqref{EQpartialLambda1} and \eqref{EQpartialLambda2} allows one to
express the derivative with respect to $\lambda$ of the extended force
density balance \eqref{EQforceDensityBalance} as
\begin{align}
  &\Sv_A(\rv)  + \cov(\beta \hat \Fv_\rmint(\rv),\hat A(\rv^N)) 
  \notag\\&\qquad
  -\nabla\chi_A(\rv) - \chi_A(\rv) \nabla \beta V_\rmext(\rv)
  = 0.
  \label{EQhyperForceDensityBalance}
\end{align}
Taking the limit $\lambda\to 0$ retains the form of
\eqr{EQhyperForceDensityBalance} and reduces the extended force
density operator therein to that of the original system, $\hat
\Fv_\rmint(\rv)= -\sum_i \delta(\rv-\rv_i)\nabla_i u(\rv^N)$.  Hence
\eqr{EQhyperForceDensityBalance} applies to general forms of
interparticle interaction potentials $u(\rv^N)$ and independently
chosen forms of the observable $\hat A(\rv^N)$.

The formally exact sum rule \eqref{EQhyperForceDensityBalance}
constitutes the locally resolved hyperforce density balance obtained
previously by Robitschko {\it et al.}~\cite{robitschko2024any}. Their
derivation of \eqr{EQhyperForceDensityBalance} rests on the both
conceptually and practically very different method of exploiting
thermal Noether invariance of the average $A$ in the standard
ensemble. Alternatively, the hyperforce density balance
\eqref{EQhyperForceDensityBalance} can also be derived from suitable
ad hoc partial integration procedures on phase space along the Yvon
theorem \cite{robitschko2024any}. Its arguably most striking, as well
as compact, derivation is that from the phase space operator methods
proposed by M\"uller {\it et al.}~\cite{mueller2024gauge} based on the
recently discovered statistical mechanical gauge invariance of
microstates \cite{mueller2024gauge, mueller2024whygauge}.

By collecting the different force contributions one can put
\eqr{EQhyperForceDensityBalance} into more compact form as
\begin{align}
  \Sv_A(\rv) +  \cov(\beta\hat\Fv(\rv), \hat A(\rv^N)) &=  0.
  \label{EQhyperForceDensityBalanceCompact}
\end{align}
Here we have introduced the total one-body force density
operator~$\hat\Fv(\rv)$ in the simple momentum-independent form
\begin{align}
  \hat\Fv(\rv) &= \hat\Fv_\rmint(\rv)-k_BT\nabla\hat\rho(\rv)
  -\hat\rho(\rv)\nabla V_\rmext(\rv),
\end{align}
which allows upon taking account of the covariance
form~\eqref{EQchiAsCovariance} of $\chi_A(\rv)$ to rewrite
\eqr{EQhyperForceDensityBalance} as
\eqr{EQhyperForceDensityBalanceCompact} \cite{robitschko2024any}.  As
Robitschko {\it et al.}~\cite{robitschko2024any} argue, the covariance
on the left hand side of \eqr{EQhyperForceDensityBalanceCompact} can
also be written as merely the mean $\langle \beta \hat\Fv(\rv)\hat
A(\rv^N)\rangle$ because the factorized contribution vanishes,
$\langle \hat \Fv(\rv)\rangle \langle \hat A(\rv^N)\rangle=0$, due to
$\langle\hat
\Fv(\rv)\rangle=\Fv_\rmint(\rv)-k_BT\nabla\rho(\rv)-\rho(\rv)\nabla\beta
V_\rmext(\rv)=0$ according to the equilibrium force density balance
\eqref{EQforceDensityBalance}.

That the hyperforce density
balance~\eqref{EQhyperForceDensityBalanceCompact} is obtained from the
present conceptually quite simple route of parametrically
differentiating the force density
balance~\eqref{EQforceDensityBalance} in the extended ensemble is a
further indication of the fundamental status of this relation (see
also the arguments given in Ref.~\cite{robitschko2024any}). We take
the apparent tight interconnection as a demonstration of the capacity
of the generalized ensemble to yield nontrivial insight. As
anticipated in our discussion of the hyperfluctuation profile
$\chi_A(\rv)$ in its covariance form \eqref{EQchiAsCovariance}, this
local measure of the coupling of fluctuations in density and in the
observable $\hat A(\rv^N)$ features prominently in the hyperforce
density balance \eqref{EQhyperForceDensityBalance} both via the
diffusion-like first term and the external force-like second term on
the left hand side of \eqr{EQhyperForceDensityBalance}.

\subsection{Hyperforces and hyperdensity functionals}
\label{SEChyperforceConnections}

The conceptual bridge between the present hyperforce concepts and the
density functional point of view of Sec.~\ref{SEChyperDFT} is provided
by linking the force density balance~\eqref{EQforceDensityBalance}
with the Euler-Lagrange equation~\eqref{EQel}. In the standard way
\cite{schmidt2022rmp} we hence build the gradient of the latter
identity, which yields upon multiplication by $\beta\rho(\rv)$ the
result
\begin{align}
  \rho(\rv)\nabla c_1(\rv;[\rho]) &= \nabla\rho(\rv) 
  + \rho(\rv)\nabla\beta V_\rmext(\rv),
  \label{EQgradEl}
\end{align}
where we have simplifed $\rho(\rv)\nabla\ln\rho(\rv)=\nabla\rho(\rv)$.
Then comparing \eqr{EQgradEl} and the force density balance
\eqref{EQforceDensityBalance} allows one, on the basis of the two
identical right hand sides, to identify:
\begin{align}
  \beta\Fv_\rmint(\rv) = \rho(\rv)\nabla c_1(\rv;[\rho]).
  \label{EQFintAsGradc1}
\end{align}
One significant consequence of \eqr{EQFintAsGradc1} is the implication
that its left hand side is elevated from a mere average
$\Fv_\rmint(\rv)=\langle \hat \Fv_\rmint(\rv) \rangle$ given by
\eqr{EQFintAsMean} to also being a density functional,
$\Fv_\rmint(\rv;[\rho])$, as given via the right hand side of
\eqr{EQFintAsGradc1}.

We recall that we still work in the extended ensemble and that hence
both sides of \eqr{EQFintAsGradc1} are understood in this way. In
order to make progress we parametrically differentiate
\eqr{EQFintAsGradc1} with respect to $\lambda$. The result is the
following identity:
\begin{align}
  &\cov(\beta \hat\Fv_\rmint(\rv),\hat A(\rv^N))
  +\Sv_A(\rv)
  \notag\\&\quad
  = \chi_A(\rv) \nabla c_1(\rv;[\rho])
  + \rho(\rv) \nabla c_A(\rv;[\rho])
  \notag\\&\quad\quad
  + \rho(\rv) \int d\rv' \chi_A(\rv') \nabla c_2(\rv,\rv';[\rho]).
  \label{EQquirky}
\end{align}
The left hand side of \eqr{EQquirky} follows directly from
\eqr{EQFintParametricDerivative}. The three terms on the right hand
side of \eqr{EQquirky} follow respectively from the right hand side of
\eqr{EQFintAsGradc1} via: i) differentiating the prefactor $\rho(\rv)$
and using \eqr{EQchiAsParametricDerivative}, ii) differentiating
$c_1(\rv;[\rho])$ parametrically upon fixing the density profile
according to \eqr{EQcAasParametricDerivative}, and iii) using the
chain rule upon monitoring the changes in the functional argument
$\rho(\rv)$ and using \eqr{EQchiAsParametricDerivative} to identify
the hyperfluctuation profile $\chi_A(\rv)$.

We can now restore the original Hamiltonian in \eqr{EQquirky} by
taking the limit $\lambda\to 0$, which in particular again sets the
interparticle force density operator to that of the original
interparticle interaction potential, $\hat \Fv_\rmint(\rv) = -\sum_i
\delta(\rv-\rv_i)\nabla_i u(\rv^N)$.  As a consistency check, we
replace the left hand side of \eqr{EQquirky} on the basis of the
hyperforce density balance~\eqref{EQhyperForceDensityBalance} and
regroup terms, which yields
\begin{align}
  &\nabla \chi_A(\rv) + \chi_A(\rv) \nabla \beta V_\rmext(\rv)
    -\chi_A(\rv) \nabla c_1(\rv;[\rho])
  \notag\\&\quad
  = \rho(\rv) \nabla c_A(\rv;[\rho])
  + \rho(\rv)\int d\rv' \chi_A(\rv') \nabla c_2(\rv,\rv';[\rho]).
  \label{EQquirky2}
\end{align}
Replacing the one-body direct correlation functional by using the
Euler-Lagrange equation \eqref{EQel} allows one to simplify the second
term on the left hand side. Upon dividing by the density profile the
result is
\begin{align}
 & \frac{\nabla \chi_A(\rv)}{\rho(\rv)} 
  + \chi_A(\rv) \frac{\nabla \ln\rho(\rv)}{\rho(\rv)}
  \notag\\ & \quad
  =\nabla c_A(\rv;[\rho])
  +\int d\rv' \chi_A(\rv') \nabla c_2(\rv,\rv';[\rho]),
  \label{EQgradHyperOZ}
\end{align}
which is identical to the gradient of the hyper-Ornstein-Zernike
relation \eqref{EQhyperOZ}.

In conclusion of this section, we find a tight integration of the
recently investigated hyperforce correlations into the hyperdensity
framework. This situation is structurally similar to the relationship
between averages of bare forces in elementary statistical physics and
in standard density functional theory, see Ref.~\cite{schmidt2022rmp}
for an extended discussion of this force point of view.

\section{Specific forms of observables}
\label{SECclassifyingObservables}

\subsection{One-body observables}
\label{SEConeBodyObervables}

The considerations presented thus far have been general in the sense
that no restrictions on the specific form of the observable $\hat
A(\rv^N)$ under consideration were applied other than that the
resulting extended ensemble is well-defined or, analogously, that
$H_A$ defined via \eqr{EQhamiltonianA} can serve as a valid
Hamiltonian and that indeed $H_A\to H$ in the limit $\lambda\to 0$.

Here we follow the SI of Ref.~\cite{sammueller2024hyperDFT} and
address specific types of $\hat A(\rv^N)$ with the aim to classify the
behaviour that arises according to different form of the dependence on
the particle configuration $\rv^N$. We first consider one-body forms
\begin{align}
  \hat A(\rv^N)&=\sum_i a_1(\rv_i), 
  \label{EQoneBodyAasSum}
\end{align}
where $a_1(\rv)$ is a given function of position~$\rv$.  Recalling the
form of the density operator \eqref{EQdensityOperator} as a sum of
Dirac distributions, we can rewrite \eqr{EQoneBodyAasSum} as
\begin{align}
  \hat A(\rv^N)=\int d\rv \hat\rho(\rv) a_1(\rv).
  \label{EQoneBodyAasIntegral}
\end{align}
Averaging \eqr{EQoneBodyAasIntegral} on both sides directly gives the
thermal mean $A$ as a density functional
\begin{align}
  A[\rho]&=\int d\rv\rho(\rv)a_1(\rv),
  \label{EQoneBodyAasDensityFunctional}
\end{align}
where we have exchanged on the right hand side the order of the
average and spatial integral and have identified $\rho(\rv)$ as the
average \eqref{EQrhoAsAverage}. From
\eqr{EQoneBodyAasDensityFunctional} it is clear that a standard
density functional treatment suffices to obtain all relevant
information as within density functional theory the density profile is
of course available.

For this simple one-body case we obtain the corresponding hyperdirect
correlation functional via insertion of
\eqr{EQoneBodyAasDensityFunctional} into
\eqr{EQcAasFunctionalDerivative}, which yields
\begin{align}
  c_A(\rv;[\rho]) &= a_1(\rv),
  \label{EQoneBodycA}
\end{align}
such that there is no density functional dependence.  The simple
result \eqref{EQoneBodycA} can serve as a toy to illustrate more
general hyperdirect correlations.

The corresponding hyperfluctuation profile $\chi_A(\rv)$ is according
to the covariance form \eqref{EQchiAsCovariance} obtained as
\begin{align}
  \chi_A(\rv)&=\int d\rv' a_1(\rv') \cov(\hat\rho(\rv),\hat\rho(\rv')) \\
  &=\int d\rv'a_1(\rv')H_2(\rv,\rv'),
  \label{EQoneBodychiA}
\end{align}
where $H_2(\rv,\rv')=\cov(\hat\rho(\rv),\hat\rho(\rv'))$ is the
standard correlation function of density fluctuations
\cite{hansen2013, evans1979}; see our discussion in the context of the
standard inhomogeneous Ornstein-Zernike equation
\eqref{EQozStandardTwoBody}.  Inserting the results
\eqref{EQoneBodycA} and \eqref{EQoneBodychiA} into the general
hyper-Ornstein-Zernike equation \eqref{EQhyperOZ} then yields this
relation upon dividing by $\beta$ in the following one-body form
\begin{align}
  \int d\rv''a_1(\rv'')  &
  \Big[   \rho(\rv)  \int d\rv' c_2(\rv,\rv')H_2(\rv',\rv'')
  \notag\\&\quad
  + \rho(\rv) \delta(\rv-\rv'')
  -H_2(\rv,\rv'') \Big]
  = 0.
  \label{EQoneBodyHyperOZ}
\end{align}
The identity \eqref{EQoneBodyHyperOZ} is straightforward to prove
directly as the term in brackets already vanishes due to the standard
two-body Ornstein-Zernike relationship \eqref{EQozStandardTwoBody}
\cite{hansen2013,schmidt2022rmp}.

In alternative reasoning we note that \eqr{EQoneBodyHyperOZ} holds for
any permissible form of $a_1(\rv)$. Hence we retain a valid identity
upon functionally differentiating with respect to~$a_1(\rv)$.
Differentiating the right hand side of \eqr{EQoneBodyHyperOZ}
trivially gives zero. Differentiating the left hand side gives the
bracketed expression, which hence necessarily needs to vanish.
Alternatively, one can argue that $a_1(\rv)$ is a mere test function
of arbitrary form and hence \eqr{EQoneBodyHyperOZ} can only hold
provided the bracketed expression vanishes. The result constitutes a
derivation of the standard two-body Ornstein-Zernike relation starting
from the hyper-Ornstein-Zernike relation~\eqref{EQoneBodyHyperOZ},
providing a demonstration of the self-consistency of the framework.

For the present one-body observables \eqref{EQoneBodyAasSum} the
following two terms that appear in \eqr{EQquirky} become identical to
each other:
\begin{align}
  \Sv_A(\rv)
  &=\rho(\rv)\nabla c_A(\rv),
  \label{EQoneBodyGradA}
\end{align}
where we recall that $\Sv_A(\rv)=\langle\sum_i
\delta(\rv-\rv_i)\nabla_i \hat A(\rv^N)\rangle$; see the corresponding
operator identity \eqref{EQhatSAdefinition}.  To prove the
relationship \eqref{EQoneBodyGradA}, we use \eqr{EQoneBodyAasSum} to
re-write the left hand side as $\langle \sum_i
\delta(\rv-\rv_i)\nabla_i a_1(\rv_i) \rangle = \rho(\rv)\nabla
a_1(\rv)=\rho(\rv)\nabla c_A(\rv)$, where in the first step we have
started from $\hat A(\rv^N)=\sum_i a_1(\rv_i)$, which gives $\nabla_i
\hat A(\rv^N)=\nabla_i a_1(\rv_i)$, and then identified the specific
form $c_A(\rv)=a_1(\rv)$ according to \eqr{EQoneBodycA}.

The result \eqr{EQoneBodycA} allows one to simplify \eqr{EQquirky} in
the following form
\begin{align}
 \int d\rv' a_1(\rv') \cov(\beta\hat\Fv_\rmint(\rv)&, \hat \rho(\rv'))
  = \chi_A(\rv)\nabla c_1(\rv;[\rho])
  \notag\\&
  +\rho(\rv) \int d\rv' \chi_A(\rv') \nabla c_2(\rv,\rv';[\rho]).
  \label{EQoneBodyQuirky}
\end{align}
On the above left hand side we have re-written
$\cov(\beta\hat\Fv_\rmint(\rv), \hat A(\rv^N))=\int d\rv' a_1(\rv')
\cov(\beta\hat\Fv_\rmint(\rv), \hat \rho(\rv'))$. We have furthermore
used \eqr{EQoneBodyGradA} to simplify.  Functional differentiation of
\eqr{EQoneBodyQuirky} with respect to $a_1(\rv'')$ yields upon taking
account of the form \eqref{EQoneBodychiA} of $\chi_A(\rv)$ the result:
\begin{align}
  \cov(\beta\hat\Fv_\rmint(\rv), \hat\rho(\rv''))
  &= H_2(\rv,\rv'')\nabla c_1(\rv;[\rho])
  \notag\\&\;
  +\rho(\rv)\int d\rv' H_2(\rv',\rv'')\nabla c_2(\rv,\rv';[\rho]).
  \label{EQfixed2}
\end{align}
Here the right hand side can be obtained as a functional derivative of
the one-body interparticle force density distribution, $-\delta
\Fv_\rmint(\rv)/\delta V_\rmext(\rv'')$, using \eqr{EQFintAsGradc1},
the functional chain rule, the identity $H_2(\rv,\rv'') =
-\delta\rho(\rv)/\delta \beta V_\rmext(\rv'')$.
The left hand side of of \eqr{EQfixed2} is the identical expression
$-\delta \Fv_\rmint(\rv)/\delta V_\rmext(\rv')$ according to the
general identity $\delta A /\delta V_\rmext(\rv) = -\beta
\cov(\hat\rho(\rv),\hat A)$ by Eckert {\it et
  al.}~\cite{eckert2023fluctuation}; see also
\eqr{EQchiAsExternalPotentialDerivative}.  We take the demonstration
that the general hyperdensity functional theory reduces to known
limits for the case of one-body observables \eqref{EQoneBodyAasSum} as
an indication for the internal consistency of our approach.

A specific example could be a simple static model countoscope
\cite{mackay2023countoscope}, where we choose $\hat A(\rv^N) = \int
d\rv a_1(\rv)\hat\rho(\rv)$ with $a_1(\rv)=\Theta(l-|\rv|)$ being an
indicator function for the spherical countoscope of diameter $2l$;
here $\Theta(\cdot)$ indicates the Heaviside unit step
function. Alternatively, one could use a version with planar symmetry,
$a_1(\rv)=\Theta(l-|x|)$.

A formally even simpler case is recovered when considering the total
number of particles $N$. Then, the framework reduces, up to a scaling
factor $\beta$, to considering the local compressibility
\cite{evans2015jpcm, evans2019pnas, coe2022prl, coe2022pre, coe2023,
  eckert2020, eckert2023fluctuation, wilding2024},
$\chi_\mu(\rv)=\beta\chi_A(\rv)$, as we demonstrate in the following.
We choose $\hat{A} = N$, which corresponds to $a_1(\rv) = 1$ in
\eqr{EQoneBodyAasSum}.  Following \eqr{EQoneBodycA} the hyperdirect
correlation functional becomes constant unity, $c_A(\rv;[\rho]) =
1$. As a consequence the hyper-Ornstein-Zernike equation
\eqref{EQhyperOZ} reduces upon multiplication by $\rho(\rv)$ to
\begin{align}
  \rho(\rv)\int
  d\rv'c_2(\rv,\rv';[\rho])\chi_\mu(\rv') + \beta\rho(\rv) =
  \chi_\mu(\rv),
\end{align}
which constitutes the exact fluctuation Ornstein-Zernike relation
\cite{eckert2020,eckert2023fluctuation} for the local compressibility
$\chi_\mu(\rv)=\beta\chi_A(\rv)$. The general functional integral
\eqref{EQmeanAviaFunctionalIntegral} reduces according to
\eqr{EQoneBodyAasDensityFunctional} to the explicit result
$A[\rho]=\int d\rv\rho(\rv)$. This clearly is the mean number of
particles, expressed as a density functional, due to $\int d\rv
\rho(\rv)=\int d\rv\langle \hat\rho(\rv)\rangle = \langle \int
d\rv\hat\rho(\rv)\rangle=\langle N \rangle=A$.

\subsection{Two-body observables}
\label{SECtwoBodyObervables}

The reduction of the hyperdensity functional framework to spatial
integration of the standard Ornstein-Zernike relation extends beyond
the above one-body forms of $\hat A(\rv^N)$. In the two-body case we
have
\begin{align}
  \hat A(\rv^N) &=  \sum_{ij}a_2(\rv_i,\rv_j),
  \label{EQtwoBodyAasSum}
\end{align}  
where the sums over $i$ and $j$ run over all particles and the
function $a_2(\rv,\rv')$ is a given two-body field. By inserting two
density operators we can re-write \eqr{EQtwoBodyAasSum} as
\begin{align}
  \hat A(\rv^N)
  &= \int d\rv d\rv' \hat\rho(\rv) \hat\rho(\rv') a_2(\rv,\rv').
  \label{EQtwoBodyAasIntegral}
\end{align}
In order to formulate the mean $A$, we build the thermal average of
\eqr{EQtwoBodyAasIntegral}.  Observing that $\langle
\hat\rho(\rv)\hat\rho(\rv')\rangle= H_2(\rv,\rv')+\rho(\rv)\rho(\rv')$
we obtain
\begin{align}
  A &= \int d\rv d\rv'
  \rho(\rv)\rho(\rv')a_2(\rv,\rv') 
  \notag\\ &\quad 
  + \int d\rv d\rv' H_2(\rv,\rv')  a_2(\rv,\rv'),
\end{align}
where the first term on the right hand side is an explicit density
functional.  Expressing the second term also as a density functional
however requires to have access to $H_2(\rv,\rv';[\rho])$ as a density
functional. Conventionally one would base this on solving the
inhomogeneous two-body Ornstein-Zernike equation
\eqref{EQozStandardTwoBody} for the specific situation at hand, which
can be numerically demanding; see
e.g.\ Refs.~\cite{tschopp2022forceDFT, sammueller2022forceDFT}.

\begin{figure*}[htb!]
  \includegraphics[width=0.95\textwidth]{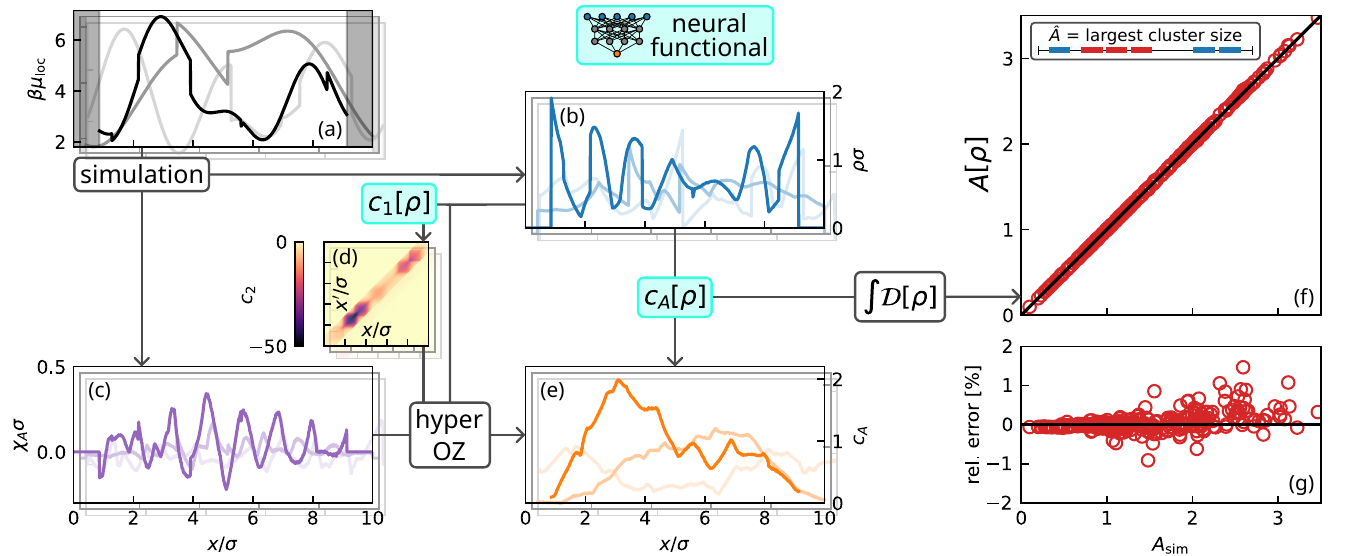}
    \caption{Overview of hyperdensity functional theory applied to the
      cluster statistics of one-dimensional hard rods of size
      $\sigma$.  The observable $\hat A$ is the size of the largest
      cluster, where two particles are bonded provided their mutual
      distance is $<1.2\sigma$.  (a)~The local chemical potential
      $\beta\mu_{\rm loc}(x)=\beta\mu-\beta V_\rmext(x)$ creates
      spatially inhomogeneous systems. Shown are representative
      examples from 512 grand canonical Monte Carlo simulations with
      both randomized values of $\beta\mu$ and forms of $\beta
      V_\rmext(x)$.  (b)~Corresponding scaled density profiles
      $\rho(x)\sigma$ sampled via \eqr{EQrhoAsAverage}.
      (c)~Corresponding scaled hyperfluctuation profiles
      $\chi_A(x)\sigma$ obtained via \eqr{EQchiAsCovariance}.
      (d)~Two-body direct correlation function $c_2(x,x';[\rho])$, as
      obtained via autodiff from $c_1(x;[\rho])$
      \cite{sammueller2023neural, sammueller2023whyNeural}.
      (e)~Hyperdirect correlation functions $c_A(x)$ obtained by
      solving the hyper-Ornstein-Zernike equation \eqref{EQhyperOZ}.
      Using the density profile as input and the simulation results
      for $c_A(x)$ as target, supervised training yields a neural
      network that represents the hyperdirect correlation functional
      $c_A(x;[\rho])$.  (f)~Predicted values $A[\rho]=\int D[\rho]
      c_A(x;[\rho])$ from functional integration according to
      \eqr{EQmeanAviaFunctionalIntegral}. For a test set of 256
      systems not encountered during training the predictions of
      $A[\rho]$ are compared against reference simulation data~$A_{\rm
        sim}$.  (g)~The relative numerical error of the predicted mean
      size $A$ of the largest cluster is smaller than $\sim 1\%$.
      Reprinted with permission from
      Ref.~\cite{sammueller2024hyperDFT}. Copyright (2024) by the
      American Physical Society.
    \label{FIG3}
   }
 \end{figure*}

\subsection{Interparticle energy as an observable}
\label{SECinterparticlePotential}

For the specific case of the considered observable being identical to
the interparticle interaction potential, $\hat A(\rv^N)=u(\rv^N)$, the
generic hyperfluctuation profile in covariance form
\eqref{EQchiAsCovariance} becomes
\begin{align}
  \chi_u(\rv)&=\cov(\hat\rho(\rv), u(\rv^N)).
  \label{EQchiu}
\end{align}
Up to a scaling factor of $1/(k_BT^2)$ this is identical to the
interparticle contribution to the local thermal susceptibility
$\chi_{T,\rmint}(\rv)=\cov(\hat\rho(\rv),u(\rv^N))/(k_BT^2)$, as
identified by Eckert {\it et al.}~\cite{eckert2023fluctuation}. Hence
$\chi_{T,\rmint}(\rv)=\chi_u(\rv)/(k_BT^2)$. These authors have proven
a contact theorem at a hard wall, $\chi_{T,\rmint}(0^+)=\langle
u(\rv^N)\rangle/(Vk_BT^2)$; see Ref.~\cite{eckert2023fluctuation} for
the derivation. From the general hypercontact
theorem~\eqref{EQhardWallContactChiTheorem} we reproduce their result
in the equivalent form $\chi_u(0^+)=\langle u(\rv^N)\rangle/V$.

In the present case the general hyperforce density balance
\eqref{EQhyperForceDensityBalance} reduces to the previously found
identity \cite{robitschko2024any}
\begin{align}
  \cov(\beta\hat \Fv(\rv), u(\rv^N)) &= \Fv_\rmint(\rv),
\end{align}
which we can re-write by using the definition of $\chi_u(\rv)$ and
re-ordering as
\begin{align}
  &\cov(\beta \hat\Fv_\rmint(\rv), u(\rv^N)) = 
  \notag\\&\quad\qquad\qquad
  \Fv_\rmint(\rv)
  +\nabla \chi_u(\rv) + \chi_u(\rv) \nabla \beta V_\rmext(\rv).
\end{align}

The general hyper-Ornstein-Zernike relation \eqref{EQhyperOZ} retains
its form as
\begin{align}
  c_u(\rv;[\rho]) &= \frac{\chi_u(\rv;[\rho])}{\rho(\rv)}
  - \int d\rv' c_2(\rv,\rv';[\rho]) \chi_u(\rv'),
\end{align}
which can be viewed as the interparticle contribution to the
fluctuation Ornstein-Zernike equation for the thermal susceptibility
$\chi_T(\rv)$ \cite{eckert2020, eckert2023fluctuation}.  We refer the
Reader to the very recent study by Kampa {\it et
  al.}~\cite{kampa2024meta}, who derived and applied
meta-Ornstein-Zernike relations on the basis of the explicit treatment
of the pair interaction potential via neural metadensity functional
dependence.

\section{Machine learning neural hyperdensity functionals}
\label{SECapplications}

\subsection{Hyperdensity functional workflow}
\label{SECworkflow}

We first lay out the principal workflow for the application of the
hyperdensity functional theory to a concrete physical problem. We
assume that all necessary functional relationships are accessible in a
concrete way. After establishing the formal procedure, we then
describe how supervised machine learning allows one to train neural
networks that provide the required functional closure for the theory.
Figure~\ref{FIG3} depicts a graphical representation of the
hyperdensity functional workflow. The relationships between the
different relevant functionals and correlation functions are
illustrated on the basis of data for the clustering in the
one-dimension hard rod model, which is taken as a prototyical case for
a nontrivial observable in Ref.~\cite{sammueller2024hyperDFT}.

We turn to the general structure.  We assume that a specific
observable $\hat A$ has been chosen as the relevant target quantity of
interest for a given many-body system, as specified by its
Hamiltonian~$H$. The extended ensemble, as described in
Sec.~\ref{SECextendedEnsemble}, requires no explicit treatment. The
aim is to study the equilibrium statistical mechanical behaviour for,
in general, spatially inhomogeneous systems at statepoints $\mu, T$,
and for given form of the external potential $V_\rmext(\rv)$.  Results
for the equilibrium density profile~$\rho(\rv)$ follow from standard
density functional minimization \eqref{EQomegaMinimal} on the basis of
the chosen excess free energy functional $F_\rmexc[\rho]$ and
corresponding one-body direct correlation functional
$c_1(\rv;[\rho])$, cf.\ their functional derivative
relationship~\eqref{EQc1definition}.  Typically one obtains the
solution of the Euler-Lagrange equation \eqref{EQel} numerically.
Then evaluating the grand potential functional $\Omega[\rho]$ gives
access to the thermodynamics. Furthermore the two-body direct
correlation functional $c_2(\rv,\rv';[\rho])$ follows from the
functional derivative relationships \eqref{EQc2fromFexc} or
\eqref{EQc2fromc1}, which gives access to the pair structure; see
e.g.\ Ref.~\cite{sammueller2024attraction} for recent work.

This standard density functional structure provides a backdrop for the
hyperfunctional application. Using the thus obtained equilibrium
density profile $\rho(\rv)$ and either evaluating the hyperdensity
functional $A[\rho]$ directly or via performing the functional
integral \eqref{EQmeanAviaFunctionalIntegral} of the one-body
hyperdirect correlation functional $c_A(\rv;[\rho])$ then gives the
desired thermal mean $A$ for the considered system.

The hyper-Ornstein-Zernike equation \eqref{EQhyperOZ} allows one to
determine the corresponding hyperfluctuation profile~$\chi_A(\rv)$. We
recall the role of $\chi_A(\rv)$ as a spatially resolved measure of
correlations of the local particle number and the value of $\hat A$
via the covariance \eqref{EQchiAsCovariance}. Here, however, we do not
have to resort to many-body sampling, as all information that is
required for the application of \eqr{EQhyperOZ} is attainable via two
functional relationships: i) evaluating the hyperdirect correlation
functional gives the concrete hyperdirect correlation function
$c_A(\rv)=c_A(\rv;[\rho])$ for the specific system at hand, and ii)
evaluating the two-body direct correlation functional,
$c_2(\rv,\rv')=c_2(\rv,\rv';[\rho])$, gives the standard two-body
direct correlation function $c_2(\rv,\rv')$, which is ready to act as
a generalized convolution kernel in the hyper-Ornstein-Zernike
equation \eqref{EQhyperOZ}. Due to $\chi_A(\rv)$ appearing both in
bare form and inside of the generalized, i.e., fully
position-dependent, spatial convolution one can solve \eqr{EQhyperOZ}
straightforwardly for $\chi_A(\rv)$, as this constitutes a system of
linear equations when using spatial discretization.

We next turn to three-dimensional hard sphere fluids and describe our
machine learning strategy for the application of the hyperdensity
functional approach to the cluster statistics, in generalization of
the one-dimensional hard rod system shown in Fig.~\ref{FIG3}.

\begin{figure*}[]
  \includegraphics[width=0.5\textwidth]{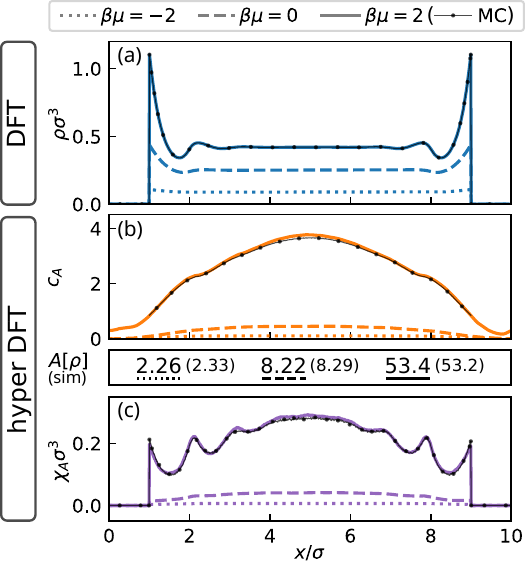}
  \includegraphics[width=0.35\textwidth]{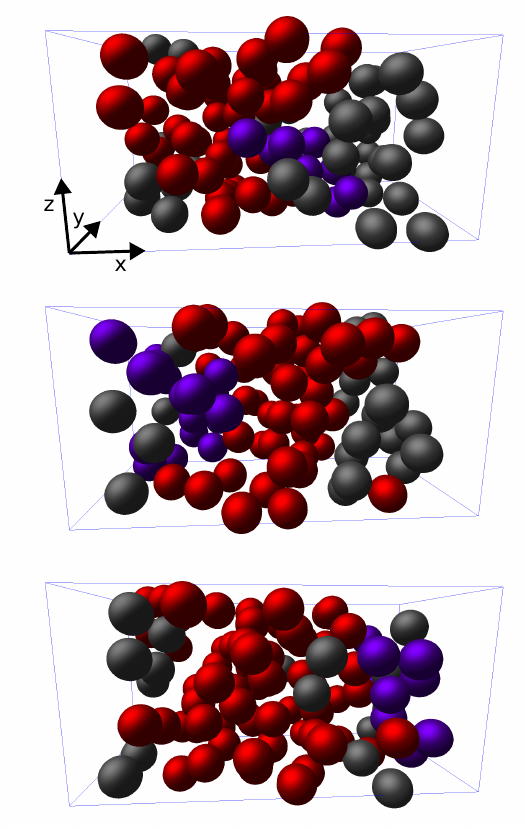}
   \caption{
   \label{FIG4} Cluster statistics of
     three-dimensional hard spheres confined between two parallel hard
     walls with distance $8\sigma$. The observable~$\hat A$ is the
     number of particles in the largest cluster.  Theoretical results
     are shown for $\beta\mu=-2$ (dotted), 0 (dashed), and 2 (solid
     lines, symbols indicate reference simulation data).
     (a)~The scaled density profile $\rho(x)\sigma^3$ as a function of
     the scaled distance $x/\sigma$ across the slit is obtained from
     numerical solution of the Euler-Lagrange equation \eqref{EQel}
     with the neural hard sphere one-body direct correlation
     functional $c_1(x;[\rho])$ of Ref.~\cite{sammueller2023neural}.
     (b)~Corresponding hyperdirect correlation functions $c_A(x)$ from
     evaluating the neural hyperdirect correlation functional
     $c_A(x;[\rho])$ at the three respective density
     profiles. Functional integration according to
     \eqr{EQmeanAviaFunctionalIntegral} gives predictions for the mean
     $A$ of the size of the largest cluster, as compared to the
     simulation reference $A=\langle \hat A \rangle$ (values in
     parenthesis).
     (c)~Hyperfluctuation profiles $\chi_A(x)$ are obtained from
     solving the hyperdirect Ornstein-Zernike relation
     \eqref{EQhyperOZ} for the three considered situations using as
     input $\rho(x)$ and the neural functionals $c_A(x;[\rho])$ and
     $c_2(x,x';[\rho])=\delta c_1(x;[\rho])/\delta\rho(x')$. The
     simulation reference is obtained via sampling
     $\chi_A(x)=\cov(\hat\rho(x), \hat A)$ according to
     \eqr{EQchiAsCovariance}.
     The three simulation snapshots (right column) show hard sphere
     configurations for $\beta\mu=2$.  The highlighted particles
     belong to the largest cluster (bright red) or to the
     second-largest cluster (dark violet). The number $\hat A$ of
     particles in the largest cluster fluctuates considerably over
     microstates.  Reprinted with permission from the SI of
     Ref.~\cite{sammueller2024hyperDFT}.  Copyright (2024) by the
     American Physical Society.}
\end{figure*}

\subsection{Training neural hyperdensity functionals}
\label{SECtraining}

\begin{figure*}
  \includegraphics[width=0.99\textwidth]{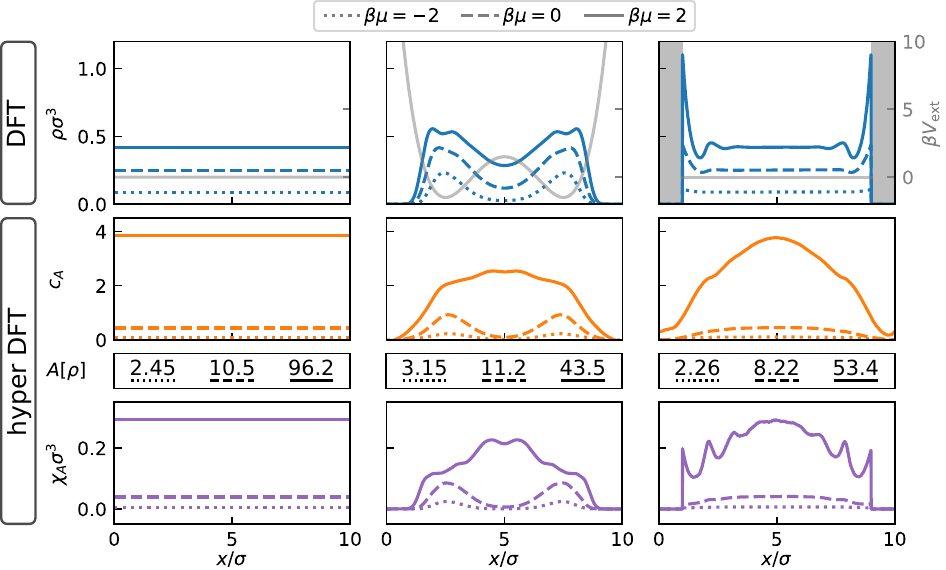}
    \caption{Hyperdensity functional results for the largest cluster
      in hard sphere fluids in bulk (left column), in a double well
      potential (middle column), and between two parallel hard walls
      (right column, reproduced from Fig.~\ref{FIG4}); see the
      corresponding scaled external potentials $\beta V_\rmext(x)$
      (gray lines in the top panels).  Shown are as a function of the
      scaled distance $x/\sigma$: the scaled density profile
      $\rho(x)\sigma^3$ (top row), the hyperdirect correlation
      function $c_A(x)$ (middle row), and the scaled hyperfluctuation
      profile $\chi_A(x)\sigma^3$ (bottom row). The scaled chemical
      potential has values $\beta\mu = -2, 0, 2$ (as indicated) and
      corresponding values of the mean size of the largest cluster
      $A[\rho]$ are given for each system. Note the striking
      structural difference between the double well potential (middle
      column) and the planar pore (right column).
    \label{FIG5}  }
\end{figure*}

Carrying out the work programme described in Sec.~\ref{SECworkflow}
requires one to have prescriptions for the excess free energy
functional $F_\rmexc[\rho]$ and for the hyperdensity functional
$A[\rho]$ to arrive at a closed theory. Alternatively and equivalently
based on functional integration, which in practice is numerically
straightforward \cite{sammueller2023neural, sammueller2023whyNeural,
  sammueller2024hyperDFT, sammueller2024attraction,
  sammueller2024pairmatching}, one can start with the one-body direct
correlation functional $c_1(\rv;[\rho])$ and the hyperdirect
correlation functional $c_A(\rv;[\rho])$. We recall their respective
relationships with $F_\rmexc[\rho]$ and $A[\rho]$ via functional
differentiation according to Eqs.~\eqref{EQc1definition} and
\eqref{EQcAasFunctionalDerivative}, as well as via functional
integration via Eqs.~\eqref{EQFexcViaFunctionalIntegral} and
\eqref{EQmeanAviaFunctionalIntegral}.

To facilitate concrete access to these functionals, we resort to
simulation-based supervised machine learning and use local learning of
one-body direct correlation functionals \cite{sammueller2023neural,
  sammueller2023whyNeural, sammueller2023neuralTutorial,
  sammueller2024pairmatching, sammueller2024attraction,
  sammueller2024hyperDFT}.  Constructing the density dependence of the
direct correlation functional $c_1(\rv;[\rho])$ follows the
methodology of Refs.~\cite{sammueller2023neural,
  sammueller2023whyNeural, sammueller2023neuralTutorial,
  sammueller2024pairmatching, sammueller2024attraction} where the
Euler-Lagrange equation~\eqref{EQel} is used to generate training data
that consists of target values $c_1(\rv)$ for given density profile
$\rho(\rv')$ within a limited range of positions $\rv'$ around the
target location $\rv$.  The application to the one-dimensional hard
rod model \cite{sammueller2023whyNeural, sammueller2023neuralTutorial}
indicates numerical performance that is in practice equivalent to that
of Percus' exact solution \cite{percus1976, robledo1981}. Functional
differentiation, as is readily accessible via automatic
differentiation \cite{sammueller2023neural, sammueller2023whyNeural,
  sammueller2023neuralTutorial, sammueller2024pairmatching,
  sammueller2024attraction, stierle2024autodiff, dijkman2024ml} yields
the two-body direct correlation functional $c_2(\rv,\rv';[\rho])$.  As
we are interested in hard sphere behaviour we re-use the trained
neural functional of Ref.~\cite{sammueller2023neural}.

Training the neural hyperdirect correlation functional
$c_A(\rv;[\rho])$ proceeds along very similar lines as training the
standard one-body direct correlation functional $c_1(\rv;[\rho])$. The
training data is acquired by grand canonical Monte Carlo simulations,
where the chemical potential $\mu$, the temperature $T$ and the shape
of the external potential $V_\rmext(\rv)$ are prescribed (in practice
in a randomized way). The density profile $\rho(\rv)$ is accessible
via sampling according to \eqr{EQrhoAsAverage} (or via more advanced
methods \cite{rotenberg2020, frenkel2023book}). For the chosen
observable $\hat A$, data for the hyperfluctuation profile
$\chi_A(\rv)$ follows from sampling the covariance
\eqref{EQchiAsCovariance}. As the two-body direct correlation
functional $c_2(\rv,\rv';[\rho])$ is known from the above standard
density functional treatment, the complete information is available to
numerically evaluate the right hand side of the hyper-Ornstein-Zernike
equation \eqref{EQhyperOZ} and hence to construct the one-body
hyperdirect correlation function $c_A(\rv)$ for the specific training
system under consideration. This data is then used as the target for
training the neural representation of the one-body hyperdirect
correlation functional $c_A(\rv;[\rho])$ using solely the density
profile $\rho(\rv')$ as an input following the lines of
Refs.~\cite{sammueller2023neural, sammueller2023whyNeural,
  sammueller2023neuralTutorial, sammueller2024pairmatching,
  sammueller2024attraction, sammueller2024hyperDFT}.  We use a simple
(fully connected) multi-layer perceptron to represent
$c_A(\rv;[\rho])$ and refer the Reader to
Ref.~\cite{sammueller2024hyperDFTzenodo} for all technical details.

We emphasize that the above laid out machine learning strategy is
heavily influenced by the successful neural functional methodology of
local learning of one-body direct correlation functionals
\cite{sammueller2023neural, sammueller2023whyNeural,
  sammueller2023neuralTutorial, sammueller2024pairmatching,
  sammueller2024attraction, sammueller2024hyperDFT}. This method
represents the classical density functional framework \cite{evans1979,
  evans1992, evans2016, hansen2013, schmidt2022rmp} in a very direct
way. Moreover, it is computationally straightforward to implement in
all its three key aspects of: data generation via simulation, training
of the neural network, and numerical application; we refer to
Ref.~\cite{sammueller2024attraction} for a demonstration of the
breadth of applicability of the resulting neural theory in the context
of gas-liquid phase separation.

This neural functional approach needs to be contrasted with brute
force machine learning, where one could envisage bypassing the density
functional relationships and working directly with the ensemble
averages in order to represent the mean $A$ via \eqr{EQmeanAasAverage}
and machine learn the relationship $V_\rmext(\rv) \to A$. The current
strategy is very different, as it resorts to machine learning only the
genuinely nontrivial hyperdirect correlation functionals and letting
the encompassing structure be informed by the theoretical physics.
This is analogous to putting the focus on $c_1(\rv;[\rho])$ in
standard DFT treatments, instead of considering other possible
mappings of interest \cite{yatsyshin2022, malpica-morales2023} or
merely mimicking the simulation procedure, i.e., learning
$V_\rmext(\rv) \to \rho(\rv)$.

\subsection{Application to cluster statistics}
\label{SECapplicationHardSpheres}

We have applied the machine learning strategy of
Sec.~\ref{SECtraining} to three-dimensional hard spheres in planar
inhomogeneous environments, choosing $\hat A$ as the size of the
largest cluster in a given microstate. Specificially, we define two
particles $i$ and $j$ as being bonded provided that their mutual
distance is below a cutoff value, which we choose as $|\rv_i-\rv_j| <
r_c=1.2\sigma$, where $\sigma$ is the hard sphere diameter. Clusters
are then defined as groups of particles that are (transitively)
bonded. Each cluster consists of a specific number of particles such
that each configuration $\rv^N$ is associated with a specific
distribution of cluster sizes. We choose the number of particles in
the largest occurring cluster as our target observable $\hat A$.

We use a fixed box size with lateral area $5\sigma\times 5\sigma$ and
an elongated $x$-direction of length $L=10\sigma$, along which the
system is spatially inhomogeneous. We first consider confinement by
two parallel hard walls represented by the scaled external potential
$\beta V_\rmext(x) = \infty$ for $x/\sigma < 1$ and $x/\sigma > 9$,
and zero otherwise. We show in Fig.~\ref{FIG4} (taken from the SI of
Ref.~\cite{sammueller2024hyperDFT}) results from the hyperdensity
functional theory compared against standalone simulation results that
provide the reference. The density profile exhibits the familiar
oscillatory behaviour, as is induced by hard sphere packing, see
Fig.~\ref{FIG4}(a) for results over a range of increasing values of
the scaled chemical potential $\beta\mu$. We have used the neural hard
sphere functional of Ref.~\cite{sammueller2023neural}, which was shown
to outperform in accuracy the White Bear Mk.~II version of fundamental
measure theory \cite{roth2002WhiteBear, roth2006WhiteBear, roth2010}.

The corresponding one-body hyperdirect correlation functions $c_A(x)$
are obtained from evaluating the neural hyperdirect correlation
functional at the respective equilibrium density profile
$\rho(x)$. Here $x$ denotes the coordinate of $\rv$ along which the
system is spatially inhomogeneous.  The spatial variation of $c_A(x)$
is much smoother than the corresponding density profile and a
pronounced broad peak is apparent at the center of the system, see
Fig.~\ref{FIG4}(b). We attribute the latter feature to the strong
effect on the size of the largest cluster that follows from having
particles near the center of the system. Crucially, via functional
integration of $c_A(x;[\rho])$, the thermal average $A$, i.e., the
mean size of the largest hard-sphere cluster, can be reproduced
faithfully, as is evident from comparison to simulation data.

The scaled hyperfluctuation profiles $\chi_A(x)\sigma^3$ shown in
Fig.~\ref{FIG4}(c) possess very pronounced oscillations with an
envelope that decays towards either wall. We recall that $\chi_A(x)$
measures the covariance \eqref{EQchiAsCovariance} of having a particle
at $x$ with the size of largest cluster in the system.

In Fig.~\ref{FIG5} we compare the clustering behaviour in the three
following situations of i) bulk fluids, where the external potential
vanishes, $V_\rmext(x) = 0$, ii) planar confinement inside of a double
well potential represented by the following form: $\beta V_\rmext(x) =
3[(x/\sigma - 5)^2 - 2.5^2]^2 / 2.5^4 - 1.5$, and iii) the above hard
wall case, which we use as a reference.  The perhaps most striking
difference between the double well and the hard wall confinement is
the overcoming of the depression that both $c_A(x)$ and $\chi_A(x)$
display at the center of the double-well system. This can be
attributed intuitively to the hump of the potential well loosing its
disrupting effect on the clustering upon increasing chemical
potential.

\section{Conclusions}
\label{SECconclusions}

In conclusion we have described in detail the recent generalization of
classical density functional theory towards the description of the
equilibrium statistical properties of virtually arbitrary observables
$\hat A(\rv^N)$. Thereby the dependence on the position coordinates
$\rv^N$ of all particles in the system can be very general and hence
can accommodate a multitude of physical quantities including complex,
intricate order parameters of interest. Our terminology parallels
Hirschfelder's hypervirial theorem \cite{hirschfelder1960} to refer to
a generalization to arbitrary observables. The hyperdensity functional
theory \cite{sammueller2024hyperDFT} is formally exact and we have
provided detailed descriptions of the underlying generalized
statistical ensemble, including the natural emergence of the
hyperfluctuation profile $\chi_A(\rv)$ as a spatially resolved measure
of fluctuations of the observable under consideration. The exact
hyper-Ornstein-Zernike equation \eqref{EQhyperOZ} relates
$\chi_A(\rv)$ to the one-body hyperdirect correlation functional
$c_A(\rv;[\rho])$ and the standard two-body direct correlation
functional $c_2(\rv,\rv';[\rho])$ plays its common role as a
generalized convolution kernel that mediates spatial correlations over
finite distances.

Our derivation of the hyper-Ornstein-Zernike
relation~\eqref{EQhyperOZ} generalizes and connects to the standard
two-body Ornstein-Zernike relation \cite{schmidt2022rmp}, the
dynamical nonequilibrium Ornstein-Zernike framework
\cite{brader2013noz, brader2014noz, schmidt2022rmp}, the local
compressibility and the local thermal susceptibility \cite{eckert2020,
  eckert2023fluctuation}, and the very recent meta-Ornstein-Zernike
framework for the explicit treatment of the form of the pair potential
\cite{kampa2024meta}.

Besides constituting the spatial inhomogeneity in the
hyper-Ornstein-Zernike equation \eqref{EQhyperOZ} and thus driving the
spatial structuring of the hyperfluctuation profile $\chi_A(\rv)$, the
hyperdirect correlation functional $c_A(\rv;[\rho])$ also offers an
avenue to express the thermal average of the observable as a genuine
density functional, $A[\rho]$. The construction of the functional
dependence involves formal functional integration, which can in
practice be numerically carried out with great efficiency provided
that the functional integrand $c_A(\rv;[\rho])$ is
available. Centering the approach around the one-body hyperdirect
correlation functional $c_A(\rv;[\rho])$ is motivated by its concrete
accessibility via simulation-based supervised machine learning. Here a
neural network is trained to act as a surrogate for the formally
defined exact functional.

The training data is provided by grand canonical Monte Carlo
simulations. Crucially, only standard techniques are thereby
necessary, without any need to explicitly consider the extended
ensemble on which the derivations are based formally. The ensemble
extension merely serves as a device to establish the required exact
functional relationships that underlie the supervised machine
learning.  Pre-processing of data involves a (standard) generalized
convolution operation with a kernel given by the two-body direct
correlation functional, which in turn is represented by a trained
neural functional \cite{sammueller2023neural, sammueller2023whyNeural,
  sammueller2023neuralTutorial}.  In practice these computational
demands are entirely manageable.

The hyperfluctuation profile $\chi_A(\rv)$ has previously emerged as a
key quantity in the context of the hyperforce correlation theory
\cite{robitschko2024any} that is based on considering thermal gauge
invariance \cite{mueller2024gauge, mueller2024whygauge} of the mean
value $A$ of the given observable. Earlier versions are the local
compressibility $\chi_\mu(\rv)$ \cite{evans2015jpcm, evans2019pnas,
  coe2022prl} and the thermal susceptibility $\chi_T(\rv)$
\cite{eckert2020, coe2022pre, eckert2023fluctuation}.  While both the
statistical mechanical Noetherian methodology
\cite{hermann2021noether, hermann2022topicalReview,
  hermann2022variance, hermann2022quantum, tschopp2022forceDFT,
  sammueller2023whatIsLiquid, hermann2023whatIsLiquid,
  robitschko2024any, sammueller2024hyperDFT} and density functional
theory itself \cite{evans1979, evans1992, evans2016, hansen2013} put
functional relationships at the center of theory construction, both
approaches are complementary and hence the respective prominent
occurrence of $\chi_A(\rv)$ is very noteworthy, see
\eqr{EQhyperForceDensityBalance}. Together with a corresponding
interparticle hyperforce contribution, the total one-body hyperforce
density correlation function is restored and ultimately related via
\eqr{EQhyperForceDensityBalanceCompact} to the mean localized phase
space gradient of the observable $\hat A(\rv^N)$ under consideration
\cite{robitschko2024any}, see the definition \eqr{EQhatSAdefinition}
of the corresponding hyperforce phase space function.

We have shown that the hyperforce density balance relationship, see
\eqr{EQhyperForceDensityBalance}, follows naturally from the extended
ensemble. The derivation is based on starting from the standard force
density balance in the extended ensemble and parametrically
differentiating with respect to the coupling parameter $\lambda$ that
tunes the strength of the generalized contribution. We take the
existence of this additional route to the hyperforce balance, besides
the derivations from Noether invariance \cite{robitschko2024any}, from
{\it ad hoc} phase space integration by parts
\cite{robitschko2024any}, and from operator methods within gauge
invariance \cite{mueller2024gauge, mueller2024whygauge} both as an
indicator for the fundamental status of this sum rule as well as for a
sign of consistent treatment of the statistical mechanics within the
hyperforce, hyperdensity, and gauge invariance frameworks.

These theoretical developments are interesting due to the fundamental
structure that they reveal to be present in the many-body statistical
physics. Furthermore, as we demonstrated, they provide a concrete
blueprint for devising machine learning schemes and in particular for
working with neural direct and hyperdirect correlation
functionals. These neural networks allow: i) immediate access to the
relevant functional dependencies, ii) efficient functional calculus
via fast numerical functional integration and via automatic functional
differentiation, as is relevant for accessing generating functionals
and higher-order correlation functions, iii) carrying out consistency
checks for the theoretical concepts and for the numerics, and iv)
gaining fresh and significantly deep insight into the correlated
many-body physics under investigation.

The relevance of the hyperdensity functional theory lies in its
combination of the rigorous aspects of classical density functional
theory, and hence the modern view of statistical mechanics of working
with functional relationships, with the wide spectrum of different
types of observables, order parameters, and general quantities of
interest that are in present-day use and often directly accessible via
many-body simulations. In such work one commonly faces the task of
finding physical meaning and structure in the simulation output. The
hyperdensity functional theory offers a formal framework for
performing this task without having to resort to costly many-body
simulations.

To provide a specific example for this strategy, we have provided a
hard wall hypercontact theorem that relates the contact value
$\chi_A(0^+)$ of the hyperfluctuation profile at a hard wall with the
mean bulk value $A_b$ of the corresponding observable per system
volume. Given this quite counter-intuitive relationship, one would
surely be hard-pressed to discover the relation by mere inspection of
simulation data. A similar contact theorem holds for Evans and
coworkers' local compressibility $\chi_\mu(\rv)$ \cite{evans2015jpcm,
  evans2019pnas, coe2022prl} as used in a multitude of interfacial
studies. As we demonstrated, the local compressibility is recovered in
the general framework when making the specific choice $\hat A=\beta
N$.

For several different classes of specific forms of the general
observable $\hat A(\rv^N)$ we have investigated the thus arising
simplifications. In particular, for one-body forms of $\hat A(\rv^N)$
all relevant information is already available within a standard
density functional treatment. Treating two-body observables within the
standard approach requires (numerical) solution of the inhomogeneous
two-body Ornstein-Zernike equation, which typically comes at
non-negligible computational expense. We have shown that the
hyperdensity approach is consistent with the standard method, while it
does not suffer from higher-order restrictions, as it entirely
operates on the one-body level. As a challenging test, we have
considered cluster statistics of hard spheres, which are inaccessible
in standard density functional treatments. As we have demonstrated,
the hyperdensity functional framework gives ready access to this
complex order parameter and its associated correlation and fluctuation
measures.

Future work could address the analytical construction of hyperdensity
functionals for specific observables. It would be interesting to see
whether concepts from fundamental measure theory \cite{rosenfeld1989,
  rosenfeld1988, roth2010} could be used in the construction of hard
sphere hyperdensity functionals. Also generalizations of fundamental
measure concepts beyond hard sphere systems acquire new relevance, in
particular the generalized weight functions for soft interactions
\cite{schmidt1999sfmt, schmidt2000sfmtMix, schmidt2000sfmtStructure,
  schmidt2007peel, schmidt2011isometric}. As was the case for the
Rosenfeld functional \cite{rosenfeld1989, rosenfeld1988, roth2010},
inspiration could come from liquid integral equation theory
\cite{hansen2013}, see Ref.~\cite{pihlajamaa2024closures} for recent
work. Also investigating relationships with the internal-energy
functional formulation \cite{schmidt2011internalEnergy} and with
functional thermodynamics \cite{anero2013} could be interesting.  We
have here worked with a fixed finite system size. Investigating the
scaling behaviour of the cluster statistics with changing system size
is a relevant topic that could possibly be addressed using the
(inverse) system size as an input to the neural functional, see
Ref.~\cite{sammueller2024attraction}. This would allow to address
multi-scale questions \cite{schmid2022editorial, brini2013multiscale,
  dellesitte2019, baptista2021}, as previously demonstrated
successfully for neural density
functionals~\cite{sammueller2023neuralTutorial, sammueller2023neural,
  sammueller2023whyNeural, sammueller2024pairmatching}.

Apart from these important conceptual points, it would be highly
interesting to use machine learning and consider the application of
neural hyperdensity functionals to a wider variety of order parameters
and systems.  This would potentially facilitate the investigation of
physical phenomena at the much increased numerical efficiency that the
neural hyperdensity functional theory delivers.  We here have provided
the full theoretical and methodological background required for
engaging in such endeavours. The generalization of the hyperdensity
functional framework to the multivariate case of several simultaneous
observables of interest will be presented elsewhere
\cite{sammueller2024multihyperDFT}.

{\bf Acknowledgments.} We thank Silas Robitschko, Sophie Hermann, and
Johanna M\"uller for useful discussions. This work is supported by the
DFG (Deutsche Forschungsgemeinschaft) under project no.~551294732.

\bigskip

{\bf Data availability statement.}  Source code, simulation data, and
neural functionals on which the present study is based are openly
available \cite{sammueller2024hyperDFTzenodo}.


\end{document}